\documentstyle[12pt,graphicx,amsfonts]{article}

\newcommand{\CN}{{\cal N}}

\newcommand{\beq}{\begin{equation}}
\newcommand{\eeq}{\end{equation}}
\newcommand{\bea}{\begin{eqnarray}}
\newcommand{\eea}{\end{eqnarray}}
\newcommand{\ena}{\end{eqnarray}}
\newcommand{\pab}{\Bar{\partial}}
\newcommand{\te}{\q}
\newcommand{\teb}{\Bar \q}
\newcommand{\F}{\Phi}
\newcommand{\Fib}{\Bar{\Phi}}
\newcommand{\Sb}{\Bar{\Sigma}}
\newcommand{\Db}{\Bar{D}}
\def\mathfrak{\bf}
\newcommand {\non}{\nonumber}
\newcommand{\bm}[1]{\mbox{\boldmath$#1$}}
\newcommand{\Ob}{\Bar{\Omega}}

\renewcommand{\(}{\left(}
\renewcommand{\)}{\right)}
\renewcommand{\[}{\left[}
\renewcommand{\]}{\right]}
\newcommand{\ggl}{\left\{}
\newcommand{\ggr}{\right\}}
\newcommand{\Dcb}{\Bar{\nabla}}
\newcommand{\Dc}{\nabla}

\newcommand{\Xis}{\breve{\Xi}}
\newcommand{\Xib}{\Bar{\Xi}}
\newcommand{\Y}{\Upsilon}
\newcommand{\Yb}{\Bar{\Upsilon}}
\newcommand{\Ys}{\breve{\Upsilon}}

\def\be{\begin{equation}}
\def\ee{\end{equation}}
\def\bea{\begin{eqnarray}}
\def\eea{\end{eqnarray}}

\def\dt#1{\on{\hbox{\bf .}}{#1}}                
\def\Dot#1{\dt{#1}}
\def\IR{\relax{\rm I\kern-.18em R}}
\def\binomial#1#2{\left(\,{\buildrel
{\raise4pt\hbox{$\displaystyle{#1}$}}\over
{\raise-6pt\hbox{$\displaystyle{#2}$}}}\,\right)}

\def\[{\lfloor{\hskip 0.35pt}\!\!\!\lceil}
\def\]{\rfloor{\hskip 0.35pt}\!\!\!\rceil}

\newcommand{\AmS}{{\protect\the\textfont2
  A\kern-.1667em\lower.5ex\hbox{M}\kern-.125emS}}

\catcode`@=11
\def\un#1{\relax\ifmmode\@@underline#1\else
        $\@@underline{\hbox{#1}}$\relax\fi}
\catcode`@=12

\def\fracm#1#2{\hbox{\large{${\frac{{#1}}{{#2}}}$}}}

\def\ad{{\kern0.5pt
                   \alpha \kern-5.05pt
\raise5.8pt\hbox{$\textstyle.$}\kern
0.5pt}}

\def\Dot#1{{\kern0.5pt
     {#1} \kern-5.05pt \raise5.8pt\hbox{$\textstyle.$}\kern
0.5pt}}

\def\a{\alpha}
\def\b{\beta}

\def\d{\delta}
\def\e{\epsilon}

\def\g{\gamma}

\def\q{\theta}

\def\z{\zeta}
\def\D{\Delta}
\def\F{\Phi}
\def\G{\Gamma}

\def\L{\Lambda}
\def\O{\Omega}

\def\S{\Sigma}

\def\bo{{\raise.15ex\hbox{\large$\Box$}}}               
\def\pa{\partial}                                       
\def\TH{{\raise.2ex\hbox{$\displaystyle \bigodot$}\mskip-4.7mu \llap H
\;}}
\def\face{{\raise.2ex\hbox{$\displaystyle \bigodot$}\mskip-2.2mu \llap
{$\ddot
        \smile$}}}                                      

\def\Bar#1{\overline{#1}}                       
\def\leftrightarrowfill{$\mathsurround=0pt \mathord\leftarrow \mkern-6mu
        \cleaders\hbox{$\mkern-2mu \mathord- \mkern-2mu$}\hfill
        \mkern-6mu \mathord\rightarrow$}
\def\dvec#1{\vbox{\ialign{##\crcr
        \leftrightarrowfill\crcr\noalign{\kern-1pt\nointerlineskip}
        $\hfil\displaystyle{#1}\hfil$\crcr}}}           
\def\dt#1{{\buildrel {\hbox{\LARGE .}} \over {#1}}}     

\def\fracm#1#2{\hbox{\large{${\frac{{#1}}{{#2}}}$}}}
\def\frac#1#2{{\textstyle{#1\over\vphantom2\smash{\raise.20ex
        \hbox{$\scriptstyle{#2}$}}}}}                   
\def\sfrac#1#2{{\vphantom1\smash{\lower.5ex\hbox{\small$#1$}}\over
        \vphantom1\smash{\raise.4ex\hbox{\small$#2$}}}} 
\def\bfrac#1#2{{\vphantom1\smash{\lower.5ex\hbox{$#1$}}\over
        \vphantom1\smash{\raise.3ex\hbox{$#2$}}}}       
\def\afrac#1#2{{\vphantom1\smash{\lower.5ex\hbox{$#1$}}\over#2}}    
\def\on#1#2{\mathop{\null#2}\limits^{#1}}               

\newskip\humongous \humongous=0pt plus 1000pt minus 1000pt
\def\caja{\mathsurround=0pt}
\def\eqalign#1{\,\vcenter{\openup2\jot \caja
        \ialign{\strut \hfil$\displaystyle{##}$&$
        \displaystyle{{}##}$\hfil\crcr#1\crcr}}\,}
\newif\ifdtup

  \def\pp{{\mathchoice
          {
              \kern 1pt%
              \raise 1pt
              \vbox{\hrule width5pt height0.4pt depth0pt
                    \kern -2pt
                    \hbox{\kern 2.3pt
                          \vrule width0.4pt height6pt depth0pt
                          }
                    \kern -2pt
                    \hrule width5pt height0.4pt depth0pt}%
                    \kern 1pt
           }
            {
              \kern 1pt%
              \raise 1pt
              \vbox{\hrule width4.3pt height0.4pt depth0pt
                    \kern -1.8pt
                    \hbox{\kern 1.95pt
                          \vrule width0.4pt height5.4pt depth0pt
                          }
                    \kern -1.8pt
                    \hrule width4.3pt height0.4pt depth0pt}%
                    \kern 1pt
            }
            {
              \kern 0.5pt%
              \raise 1pt
              \vbox{\hrule width4.0pt height0.3pt depth0pt
                    \kern -1.9pt  
                    \hbox{\kern 1.85pt
                          \vrule width0.3pt height5.7pt depth0pt
                          }
                    \kern -1.9pt
                    \hrule width4.0pt height0.3pt depth0pt}%
                    \kern 0.5pt
            }
            {
              \kern 0.5pt%
              \raise 1pt
              \vbox{\hrule width3.6pt height0.3pt depth0pt
                    \kern -1.5pt
                    \hbox{\kern 1.65pt
                          \vrule width0.3pt height4.5pt depth0pt
                          }
                    \kern -1.5pt
                    \hrule width3.6pt height0.3pt depth0pt}%
                    \kern 0.5pt
            }
        }}

  \def\mm{{\mathchoice
                       {
                             \kern 1pt
               \raise 1pt    \vbox{\hrule width5pt height0.4pt depth0pt
                                  \kern 2pt
                                  \hrule width5pt height0.4pt depth0pt}
                             \kern 1pt}
                       {
                            \kern 1pt
               \raise 1pt \vbox{\hrule width4.3pt height0.4pt depth0pt
                                  \kern 1.8pt
                                  \hrule width4.3pt height0.4pt depth0pt}
                             \kern 1pt}
                       {
                            \kern 0.5pt
               \raise 1pt
                            \vbox{\hrule width4.0pt height0.3pt depth0pt
                                  \kern 1.9pt
                                  \hrule width4.0pt height0.3pt depth0pt}
                            \kern 1pt}
                       {
                           \kern 0.5pt
             \raise 1pt  \vbox{\hrule width3.6pt height0.3pt depth0pt
                                  \kern 1.5pt
                                  \hrule width3.6pt height0.3pt depth0pt}
                           \kern 0.5pt}
                       }}

\def\pd{{\kern0.5pt
                   + \kern-5.05pt \raise5.8pt\hbox{$\textstyle.$}\kern
0.5pt}}

\def\pmd{{\kern0.5pt
                  \pm \kern-5.05pt \raise6.3pt\hbox{$\textstyle.$}\kern1.5pt}}

\def\md{{\mathchoice
   {
      {{\kern 1pt - \kern-6.2pt \raise5pt\hbox{$\textstyle.$}\kern 1pt}}}
    {
      {{\kern 1pt - \kern-6.2pt \raise5pt\hbox{$\textstyle.$}\kern 1pt}}}
    {
      {\kern0.5pt - \kern-5.05pt \raise3.4pt\hbox{$\textstyle.$}\kern0.5pt}}
    {
      {\kern0.5pt - \kern-5.05pt \raise3.4pt\hbox{$\textstyle.$}\kern0.5pt}}}}

\def\ad{{\dot{\alpha}}}
\def\bd{{\dot{\beta}}}

\def\pp{{\mathchoice
          {
              \kern 1pt%
              \raise 1pt
              \vbox{\hrule width5pt height0.4pt depth0pt
                    \kern -2pt
                    \hbox{\kern 2.3pt
                          \vrule width0.4pt height6pt depth0pt
                          }
                    \kern -2pt
                    \hrule width5pt height0.4pt depth0pt}%
                    \kern 1pt
           }
            {
              \kern 1pt%
              \raise 1pt
              \vbox{\hrule width4.3pt height0.4pt depth0pt
                    \kern -1.8pt
                    \hbox{\kern 1.95pt
                          \vrule width0.4pt height5.4pt depth0pt
                          }
                    \kern -1.8pt
                    \hrule width4.3pt height0.4pt depth0pt}%
                    \kern 1pt
            }
            {
              \kern 0.5pt%
              \raise 1pt
              \vbox{\hrule width4.0pt height0.3pt depth0pt
                    \kern -1.9pt  
                    \hbox{\kern 1.85pt
                          \vrule width0.3pt height5.7pt depth0pt
                          }
                    \kern -1.9pt
                    \hrule width4.0pt height0.3pt depth0pt}%
                    \kern 0.5pt
            }
            {
              \kern 0.5pt%
              \raise 1pt
              \vbox{\hrule width3.6pt height0.3pt depth0pt
                    \kern -1.5pt
                    \hbox{\kern 1.65pt
                          \vrule width0.3pt height4.5pt depth0pt
                          }
                    \kern -1.5pt
                    \hrule width3.6pt height0.3pt depth0pt}%
                    \kern 0.5pt
            }
        }}

  \def\mm{{\mathchoice
                       {
                             \kern 1pt
               \raise 1pt    \vbox{\hrule width5pt height0.4pt depth0pt
                                  \kern 2pt
                                  \hrule width5pt height0.4pt depth0pt}
                             \kern 1pt}
                       {
                            \kern 1pt
               \raise 1pt \vbox{\hrule width4.3pt height0.4pt depth0pt
                                  \kern 1.8pt
                                  \hrule width4.3pt height0.4pt depth0pt}
                             \kern 1pt}
                       {
                            \kern 0.5pt
               \raise 1pt
                            \vbox{\hrule width4.0pt height0.3pt depth0pt
                                  \kern 1.9pt
                                  \hrule width4.0pt height0.3pt depth0pt}
                            \kern 1pt}
                       {
                           \kern 0.5pt
             \raise 1pt  \vbox{\hrule width3.6pt height0.3pt depth0pt
                                  \kern 1.5pt
                                  \hrule width3.6pt height0.3pt depth0pt}
                           \kern 0.5pt}
                       }}

\def\pd{{\kern0.5pt
                   + \kern-5.05pt \raise5.8pt\hbox{$\textstyle.$}\kern
0.5pt}}

\def\pmd{{\kern0.5pt
                  \pm \kern-5.05pt \raise6.3pt\hbox{$\textstyle.$}\kern1.5pt}}

\def\md{{\mathchoice
   {
      {{\kern 1pt - \kern-6.2pt \raise5pt\hbox{$\textstyle.$}\kern 1pt}}}
    {
      {{\kern 1pt - \kern-6.2pt \raise5pt\hbox{$\textstyle.$}\kern 1pt}}}
    {
      {\kern0.5pt - \kern-5.05pt \raise3.4pt\hbox{$\textstyle.$}\kern0.5pt}}
    {
      {\kern0.5pt - \kern-5.05pt \raise3.4pt\hbox{$\textstyle.$}\kern0.5pt}}}}

\def\dslash{\not{\hbox{\kern-2pt $\partial$}}}
\def\Dslash{\not{\hbox{\kern-4pt $D$}}}
\def\pslash{\not{\hbox{\kern-2.3pt $p$}}}
 \newtoks\slashfraction
 \slashfraction={.13}
 \def\slash#1{\setbox0\hbox{$ #1 $}
 \setbox0\hbox to \the\slashfraction\wd0{\hss \box0}/\box0 }

\font\ro=cmsy10                          
\def\kcr{{\hbox{\ro \char'170}}}                
\def\ktl{{\hbox{\ro \char'170}}}        
\def\ktr{{\hbox{\ro \char'170}}}        
\def\kbl{{\hbox{\ro \char'170}}}        
\def\kbr{{\hbox{\ro \char'170}}}        

\def\plpl{\raise-2pt\hbox{$\raise3pt\hbox{$_+$}\hskip-6.67pt\raise0.0pt
\hbox{$^+$}\hskip 0.01pt$}}
\def\mimi{\raise-2pt\hbox{$\raise3pt\hbox{$_-$}\hskip-6.67pt\raise0.0pt
\hbox{$^-$}\hskip 0.01pt$}}

\def\bo{{\raise.15ex\hbox{\large$\Box$}}}               
\def\pa{\partial}                                       
\def\TH{{\raise.2ex\hbox{$\displaystyle \bigodot$}\mskip-4.7mu \llap H \;}}
\def\face{{\raise.2ex\hbox{$\displaystyle \bigodot$}\mskip-2.2mu \llap {$\ddot
        \smile$}}}                                      

\def\Bar#1{\overline{#1}}                       
\def\leftrightarrowfill{$\mathsurround=0pt \mathord\leftarrow \mkern-6mu
        \cleaders\hbox{$\mkern-2mu \mathord- \mkern-2mu$}\hfill
        \mkern-6mu \mathord\rightarrow$}
\def\dvec#1{\vbox{\ialign{##\crcr
        \leftrightarrowfill\crcr\noalign{\kern-1pt\nointerlineskip}
        $\hfil\displaystyle{#1}\hfil$\crcr}}}           
\def\dt#1{{\buildrel {\hbox{\LARGE .}} \over {#1}}}     

\def\fracm#1#2{\hbox{\large{${\frac{{#1}}{{#2}}}$}}}
\def\frac#1#2{{\textstyle{#1\over\vphantom2\smash{\raise.20ex
        \hbox{$\scriptstyle{#2}$}}}}}                   
\def\sfrac#1#2{{\vphantom1\smash{\lower.5ex\hbox{\small$#1$}}\over
        \vphantom1\smash{\raise.4ex\hbox{\small$#2$}}}} 
\def\bfrac#1#2{{\vphantom1\smash{\lower.5ex\hbox{$#1$}}\over
        \vphantom1\smash{\raise.3ex\hbox{$#2$}}}}       
\def\afrac#1#2{{\vphantom1\smash{\lower.5ex\hbox{$#1$}}\over#2}}    
\def\on#1#2{\mathop{\null#2}\limits^{#1}}               

\parskip=\medskipamount                 
\thicklines                         

\def\oldheadpic{                                
        \setlength{\unitlength}{.4mm}
        \thinlines
        \par
        \begin{picture}(349,16)
        \put(325,16){\line(1,0){4}}
        \put(330,16){\line(1,0){4}}
        \put(340,16){\line(1,0){4}}
        \put(335,0){\line(1,0){4}}
        \put(340,0){\line(1,0){4}}
        \put(345,0){\line(1,0){4}}
        \put(329,0){\line(0,1){16}}
        \put(330,0){\line(0,1){16}}
        \put(339,0){\line(0,1){16}}
        \put(340,0){\line(0,1){16}}
        \put(344,0){\line(0,1){16}}
        \put(345,0){\line(0,1){16}}
        \put(329,16){\oval(8,32)[bl]}
        \put(330,16){\oval(8,32)[br]}
        \put(339,0){\oval(8,32)[tl]}
        \put(345,0){\oval(8,32)[tr]}
        \end{picture}
        \par
        \thicklines
        \vskip.2in}
\def\oldtitle#1#2#3#4{\oldheadpic\begin{center}\vglue.5in{\large\bf #1}\\[.6in]
        {#2}\\[.1in] {\it Department of Physics and Astronomy}\\
        {\it University of Maryland, College Park, MD 20742}\\[.6in]
        Physics Publication \#{#3}\\ {#4}\\[1.5in] {\bf ABSTRACT}\\[.1in]
        \end{center} \begin{quotation}}                 
\def\oldTitle#1#2#3#4#5#6#7{\oldheadpic\begin{center} \vglue .4in
        {\large\bf #1}\\[.4in]
        {#2}\\[.1in] {\it Department of Physics and Astronomy}\\
        {\it University of Maryland, College Park, MD 20742}\\[.1in]
        {#3}\\[.1in] {\it {#4}}\\ {\it {#5}}\\[.4in]
        Physics Publication \#{#6}\\ {#7}\\[.5in] {\bf ABSTRACT}\\[.1in]
        \end{center} \begin{quotation}}                 
\def\border{                                            
        \setlength{\unitlength}{1mm}
        \newcount\xco
        \newcount\yco
        \xco=-21
        \yco=12
        \begin{picture}(140,0)
        \put(\xco,\yco){$\ktl$}
        \advance\yco by-1
        {\loop
        \put(\xco,\yco){$\kcr$}
        \advance\yco by-2
        \ifnum\yco>-240
        \repeat
        \put(\xco,\yco){$\kbl$}}
        \xco=158
        \yco=12
        \put(\xco,\yco){$\ktr$}
        \advance\yco by-1
        {\loop
        \put(\xco,\yco){$\kcr$}
        \advance\yco by-2
        \ifnum\yco>-240
        \repeat
        \put(\xco,\yco){$\kbr$}}
        \put(-20,13){\tiny **University of Maryland * Center for String and
         Particle  Theory* Physics Department***University of Maryland *Center
        for String and Particle  Theory** }
        \put(-20,-241.5){\tiny **University of Maryland * Center for String and
         Particle  Theory* Physics Department***University of Maryland *Center
        for String and Particle  Theory** }
        \end{picture}
        \par\vskip-8mm}
\def\bordero{                                           
        \setlength{\unitlength}{1mm}
        \newcount\xco
        \newcount\yco
        \xco=-31
        \yco=12
        \begin{picture}(140,0)
        \put(\xco,\yco){$\ktl$}
        \advance\yco by-1
        {\loop
        \put(\xco,\yco){$\kclr$}
        \advance\yco by-2
        \ifnum\yco>-240
        \repeat
        \put(\xco,\yco){$\kbl$}}
        \xco=151
        \yco=12
        \put(\xco,\yco){$\ktr$}
        \advance\yco by-1
        {\loop
        \put(\xco,\yco){$\kcr$}
        \advance\yco by-2
        \ifnum\yco>-240
        \repeat
        \put(\xco,\yco){$\kbr$}}
        \put(-20,12){\ooo bacdefghidfghghdhededbihdgdfdfhhdheidhdhebaaahjhhdahba

hgdedge
   hgfdiehhgdigicba}
        \put(-20,-241.5){\ooo ababaighefdbfghgeahgdfgafagihdidihiidhiagfedhadbfd

ecdcdfa
   gdcbhaddhbgfchbgfdacfediacbabab}
        \end{picture}
        \par\vskip-8mm}
\def\headpic{                                           
        \indent
        \setlength{\unitlength}{.4mm}
        \thinlines
        \par
        \begin{picture}(29,16)
        \put(165,16){\line(1,0){4}}
        \put(170,16){\line(1,0){4}}
        \put(180,16){\line(1,0){4}}
        \put(175,0){\line(1,0){4}}
        \put(180,0){\line(1,0){4}}
        \put(185,0){\line(1,0){4}}
        \put(169,0){\line(0,1){16}}
        \put(170,0){\line(0,1){16}}
        \put(179,0){\line(0,1){16}}
        \put(180,0){\line(0,1){16}}
        \put(184,0){\line(0,1){16}}
        \put(185,0){\line(0,1){16}}
        \put(169,16){\oval(8,32)[bl]}
        \put(170,16){\oval(8,32)[br]}
        \put(179,0){\oval(8,32)[tl]}
        \put(185,0){\oval(8,32)[tr]}
        \end{picture}
        \par\vskip-6.5mm
        \thicklines}
\def\title#1#2#3#4{\border\headpic {\hbox to\hsize{#4 \hfill UMDEPP #3}}\par
        \begin{center} \vglue .5in {\large\bf #1}\\[.6in]
        {#2}\\[.1in] {\it Department of Physics and Astronomy}\\
        {\it University of Maryland, College Park, MD 20742}\\[1.5in]
        {\bf ABSTRACT}\\[.1in] \end{center} \begin{quotation}}  
\def\Title#1#2#3#4#5#6#7{\border\headpic
        {\hbox to\hsize{#7 \hfill UMDEPP #6}}\par
        \begin{center} \vglue .4in {\large\bf #1}\\[.4in]
        {#2}\\[.1in] {\it Department of Physics and Astronomy}\\
        {\it University of Maryland, College Park, MD 20742}\\[.1in]
        {#3}\\[.1in] {\it {#4}}\\ {\it {#5}}\\[.5in] {\bf ABSTRACT}\\[.1in]
        \end{center} \begin{quotation}}                 
\def\endtitle{\end{quotation}\newpage}                  

\def\qd{{\kern0.5pt
                   q \kern-5.05pt \raise5.8pt\hbox{$\textstyle.$}\kern
0.5pt}}

\begin{document}

\def\dt#1{\on{\hbox{\bf .}}{#1}}                
\def\Dot#1{\dt{#1}}

\def\gfrac#1#2{\frac {\scriptstyle{#1}}
        {\mbox{\raisebox{-.6ex}{$\scriptstyle{#2}$}}}}
\def\gg{{\hbox{\sc g}}}
\border\headpic {\hbox to\hsize{April 2006 \hfill
{UMDEPP 06-003}}}
\par
{$~$ \hfill
{Bicocca--FT--06--6}}
\par
~~~
{$~$ \hfill {hep-th/0604042}}
\par

\setlength{\oddsidemargin}{0.3in}
\setlength{\evensidemargin}{-0.3in}
\begin{center}
\vglue .10in
{\large\bf 6D Supersymmetric Nonlinear Sigma-Models in\\
4D, $\bm {\cal N}$ $\bm =$ 1 Superspace}

S.\, James Gates, Jr.${}^\dag$\footnote{gatess@wam.umd.edu}, 
Silvia Penati${}^\star$\footnote{Silvia.Penati@mib.infn.it} and
Gabriele Tartaglino-Mazzucchelli${}^\star$\footnote{Gabriele.Tartaglino@mib.infn.it}
\\[0.3in]
${}^\dag${\it Center for String and Particle Theory\\
Department of Physics, University of Maryland\\
College Park, MD 20742-4111 USA}\\[0.1in]
{\it {and}}\\[0.1in]
${}^\star${\it Dipartimento di Fisica, Universit\`a degli studi
Milano-Bicocca\\and INFN, Sezione di Milano-Bicocca, piazza della Scienza 3,
I-20126 Milano, Italy}\\[0.6in]

{\bf ABSTRACT}\\[.01in]
\end{center}
\begin{quotation}
{Using 4D, $\cal N$ $=$ 1 superfield techniques, a discussion of the 6D 
sigma-model possessing simple supersymmetry is given.  Two such approaches are 
described.
Foremost it is shown that the simplest and most transparent description arises 
by use of a doublet of chiral scalar superfields for each 6D hypermultiplet.  
A second description that is most directly related to projective superspace is
also presented.
The latter necessarily implies the use of one chiral superfield and one 
nonminimal scalar superfield for each 6D hypermultiplet.  A separate study of 
models of this class, outside the context of projective superspace, is also 
undertaken.}

\endtitle

\setlength{\oddsidemargin}{0.3in}
\setlength{\evensidemargin}{-0.3in}

\setcounter{equation}{0}
\section{Introduction}

The topic of six dimensional supersymmetrical sigma-models 
\cite{SUSYKHSTK6d,SieTow} is curiously one that has hardly been explored in the
literature.   Certainly one possible explanation
for this is the expectation that no fundamentally new features will emerge. For
example, since by reduction to 4D they become $\cal N$ = 2 models, 
the already extensive literature on the latter must surely constitute
an indirect study of these models and has already illustrated all
structures of the 6D theories.  However, this raises questions that always 
occur
when discussions of compactifications are present in supersymmetrical theories.
Are there features of the compactified theories that only occur in the lower 
dimension?
How are the features that only permitted in the 6D theory to be 
disentangled from those that are present only in the compactified theory?  
Moreover, with the topic of `little strings' 
\cite{littleString} having been discovered, one would also prefer a
study of 6D nonlinear sigma-model theory in an effort to find whether there are
features of the former that are encoded in the structure of the latter.  
Finally, these
studies of 6D models in terms of 4D,  $\cal N$ = 1 (or more generally higher D)
models \cite{siegel}--\cite{ArkaniHamedTB} 
opens up an arena for the study of the corresponding realizations of
superconformal symmetry, supergravity and perhaps most fascinating of all, 
superstring/M-theory.

In previous work 
\cite{0508187} 
we have probed the structure of the 6D 
hypermultiplet as viewed by the tool of a formulation that only realizes the 
full 6D Lorentz group
fully on-shell but permits the realization of 4D, $\cal N$ = 1 supersymmetry 
off-shell.  Thus, the present work naturally follows onto this previous set of
investigations.  A summary of this work follows.

In the second chapter, the formulation of this class of models in terms of
pair of chiral multiplets (CC formulation) is given.  It is shown how the
condition of on-shell Lorentz invariance naturally leads to the condition that
the geometry of the nonlinear sigma-models must be that of a hyper-K\"ahler 
manifold \cite{GaumeFreedman}.  
The determinant of the hyper-K\"ahler metric 
is equal to the square modulus of the determinant of the exterior 
derivative of a holomorphic one-form
which is related, in our 4D, $\CN=1$ superspace fomulation, 
to the extra-dimensions.
This condition results to be equivalent to the Monge-Amp\`ere equation 
and implies Ricci flatness.
The triplet of complex structures that possess a quaternionic algebra is
identified and related to the exterior derivative of the holomorphic 
one-form.
With a correct definition of how to obtain the 6D component fields from the 4D
ones, the on-shell action is found to take the expected form: Kinetic
energies for the spin-zero and spin-1/2 states together with a quartic
fermionic interaction that involves the Riemann tensor for the manifold
geometry.

In the third chapter, an exploration of the origin of such models arising
from projective superspace 
\cite{ProjectiveSuperspace}--\cite{GrunLind}, \cite{0508187}
is undertaken.  `Projectivized' superderivatives
are defined in the usual manner.  This is followed by a review of the polar
formulation of hypermultiplets and the discussion of sigma-model actions
that can be introduced for these. 
As an example of the general structure of these 6D sigma-models we 
consider the particular case of tangent bundles of K\"ahler manifolds.  
Although no explicit results are given
for the $O(2n)$ 6D ${\cal N}$ $=$ $(1,\,0)$ multiplets, it is noted that the 
extension to the 6D arena is possible.

In the fourth chapter, we analyze the very difficult problem of deriving the
geometry that arises in the case of directly using the CNM 
(chiral/non-minimal) 
\cite{DeoGates}--\cite{9803230}
formulation without the starting point of projective superspace.  The starting
point for this mimics the techniques used in chapter two but includes now
the complication to allow {\it {both}} chiral and complex linear superfields
\cite{SUPERSPACE,QuantizedNM} 
(i.e. non-minimal scalar multiplets) {\it {ab initio}} in the analysis.  
It is noted
that whenever the number of nonminimal multiplets is less than the number
of chiral multiplet, a subsector of the theory must take the form given in
chapter two.  Full expressions for the bosonic terms in the action, prior
to removal of auxiliary fields are given.  Imposing 6D Lorentz invariance,
imposes a condition on the generalized potential in the model that is very
similar to that found in the pure CC case.  However, no simple solution
to the general case of this system are obtainable by our present methods. 
An explicit solution is presented in a special case where an
explicit proof is obtained that CNM geometry is a hyper-K\" ahler one.

In the fifth chapter, a discussion of the duality between the
6D CC and CNM formulations is undertaken.  Once again the analysis of
the general case is hampered by the sheer complexity of the problem.  
Subject to a special choice of a Darboux sympletic atlas, the results indicate
no obstructions to carrying out such duality maps.  

In the sixth chapter, there is presented an indirect study of the CNM sigma-
models via the use of duality with respect to CC models.  This allows
a direct inference of the constraints of the CNM model by using the
duality of their correspondence to objects that occur in the CC approach.

We include a chapter with our conclusions and include two appendices.
The first appendix is used to state the conventions of the paper.
The second contains explicit calculations of the actions that involve the
CNM formulation to obtain component level results.

\setcounter{equation}{0}
\section{6D, $\CN=(1,0)$ CC sigma-models}

We formulate six--dimensional nonlinear sigma--models using a
formalism which keeps four dimensional $\CN=1$ 
supersymmetry\footnote{We use the
conventions of \cite{SUPERSPACE} and \cite{0508187}.} manifest. 

The 6D, $\CN=(1,0)$ hypermultiplet can be described
in terms of two chiral multiplets \cite{MaSaSi,ArkaniHamedTB} (CC formulation) 
or one chiral multiplet and one complex linear multiplet \cite{0508187} 
(CNM formulation). We start considering the CC formulation.

The action which describes the free dynamics of a 6D, $\CN=(1,0)$ CC 
hypermultiplet \cite{MaSaSi,ArkaniHamedTB,0508187} is
\be
\eqalign{ {~~~~~~}
S_{CC}~=~&\int d^6x\, d^4\q \, \Big[~\Bar{\F}_+ \, \F_+ ~+~
\Bar{\F}_- \, \F_-~\Big] ~+~
\int d^6x \, d^2\te \Big[~\F_+\, \pa \,\F_-~\Big] \cr
&+~\int d^6x\, d^2\teb \Big[~\Bar{\F}_+\,\pab \,\Bar{\F}_-~\Big]~~~,}
\label{S0CC}
\eeq
where 
\be
\eqalign{ &z\equiv{1\over 2}(x_4+ix_5)~~~~~,~~~~~
\pa\equiv{\pa\over \pa z}=\pa_4-i\pa_5~~~~;~~~~  \cr
&\Bar{z}\equiv{1\over 2}(x_4-ix_5)~~~~~,~~~~~ \pab\equiv{\pa\over
\pa \Bar{z}}=\pa_4+i\pa_5~~~~.~~~~ }
\label{defZbZ}
\ee
The action (\ref{S0CC}) is explicitly invariant under
Sl$(2,\mathbb{C})\times$U$(1)\simeq\,$
SO$(1,3)\times\,$SO$(2)\subset\,$SO$(1,5)$, a proper subgroup of the 6D Lorentz
group, and it has off--shell 4D, $\CN=1$ SUSY.
The ${\rm U}(1)\simeq {\rm SO}(2)$ is the subgroup of rotations on
the $(4,5)$-plane in 6D Minkowski space and acts as phase transformations on 
$\pa\to e^{i\phi}\pa$, $\pab\to e^{-i\phi}\pab$. 
The (anti)chiral superfields of the hypermultiplet
are assumed to be neutral under the ${\rm U}(1)$ subgroup since the bosonic 
physical fields $A_\pm=\F_\pm|$ and $\Bar{A}_\pm=\Fib_\pm|$ must be neutral
(i.e. scalars with respect to the 6D Lorentz group).
From the invariance of the holomorphic\footnote{We use '
holomorphic terms' instead of 'superpotential terms' since they lead to
the appearance
\\$~~~~~~$
of derivatives of the propagating bosons.} terms in 
(\ref{S0CC}), it follows that the grassmannian differentials transform as
$d\te_\a\to e^{-{i\over 2}\phi}d\te_\a$, $d\teb_\ad\to 
e^{{i\over 2}\phi}d\teb_\ad$.

Once integrated out, the auxiliary fields in (\ref{S0CC}) lead to a resulting 
action which has linearly realized 6D Lorentz invariance and is on--shell
6D, $\CN=(1,0)$  supersymmetric.

We now extend this analysis to 6D nonlinear sigma--models and find restrictions
on the target space geometry induced by the request for the model to be 6D 
covariant and supersymmetric. 

We start generalizing the action (\ref{S0CC}) to a system of $n$ decoupled 
CC hypermultiplets describing a flat complex $2n$--dimensional 
target space. Defining $\Psi^a=(\F_+^I,\F_-^i)$ we write
\beq
S = \int d^6x \Bigg[\int d^4\te\, \Bar{\Psi}^{\Bar{a}}\d_{\Bar{a}b}\Psi^b+
{1\over 2}\int d^2\te\,\Psi^a\,\O_{ab}\,\pa\,\Psi^b+
{1\over 2}\int d^2\teb\,\Bar{\Psi}^{\Bar{a}}\,\Ob_{\Bar{a}\Bar{b}}\,\pab\,
\Bar{\Psi}^{\Bar{b}}~\Bigg]~~~,~~~
\label{flatSigma6D}
\eeq
where 
\beq
\d_{\Bar{a}b}=\begin{footnotesize}\(\begin{array}{cc}\d_{\Bar{I}J}&0\\0 &
\d_{\Bar{i}j}\end{array}\)\end{footnotesize}
\qquad \qquad 
\O_{ab}=\Ob_{\Bar{a}\Bar{b}}=
\begin{footnotesize}\(\begin{array}{cc}0&
\d_{Ij}\\-\d_{iJ}& 0\end{array}\)\end{footnotesize} \, .
\eeq
To extend non--trivially the action (\ref{flatSigma6D}) to a curved target 
space we make the following ansatz
\beq
\int d^6x\Bigg[
\int d^4\te~K\(\Psi^a,\Bar{\Psi}^{\Bar{a}}\)+\int d^2\te~
Q_a\Big(\Psi^b\Big)\pa\,\Psi^a+ \int d^2\teb~
\Bar{Q}_{\Bar{a}}\Big(\Bar{\Psi}^{\Bar{b}}\Big)\pab\,
\Bar{\Psi}^{\Bar{a}}~\Bigg]~~~.
\label{kahler6DCC}
\eeq
Here the functions $Q_a$ ($\Bar{Q}_{\Bar{a}}$) are (anti)holomorphic
in the (anti)chiral superfields $\Psi^a$ ($\Bar{\Psi}^{\Bar{a}}$).
The expression (\ref{kahler6DCC}) is the most general ansatz for an action 
local in the physical 
fields which generalizes (\ref{flatSigma6D}) and still has the 
off--shell symmetries of the flat case, i.e. 4D SUSY and the 
${\rm Sl}(2,{\mathbb{C}})\times {\rm U}(1)$ invariance.

A feature of note regarding (\ref{kahler6DCC}) is the appearance of
$Q_a (\Psi^b)$
in the extra--dimensions derivatives holomorphic term.
This quantity has an interpretation as the connection of
a U$(1)$-bundle.  This U$(1)$-bundle is not necessarily related to the
one that is part  of ${\rm Sl}(2,{\mathbb{C}})\times {\rm U}(1)$ invariance.
In fact, the U$(1)$-bundle for which $Q_a (\Psi^b)$ is the connection
is a bundle defined over the manifold.  The fact that $Q_a (\Psi^b)$ 
appears as it does in (\ref{kahler6DCC}) implies that it is ambiguous with 
respect the gauge transformation
\beq
Q_a  (\Psi^b) ~\to~ Q_a (\Psi^b) ~+~ {{\pa ~~~} \over {\pa \Psi^a }} {\cal T}
(\Psi^b)
~~~,
\label{Qgauge}
\eeq
since the purely holomorphic terms are only changed by surface terms with 
regard to this redefinition.  This invariance will be seen at the level of the 
action by the result that this U$(1)$-bundle connection will only appear in 
quantities via its exterior derivative.

It is important to note that the rigid ${\rm U}(1)$ invariance and as well
the local manifold U$(1)$-bundle invariance, both fix the form of the
the latter two terms in (\ref{kahler6DCC}) and exclude the possibility to have
terms like $\int d^2\te\widetilde{Q}_a\pab\,\Psi^a+
\int d^2\teb\Bar{\widetilde{Q}}_{\Bar{a}}\,\pa\,\Bar{\Psi}^{\Bar{a}}$. In 
analogy with the flat space \cite{0508187}, such contributions would be the 
only possible terms admitted if we were to impose opposite ${\rm U}(1)$ phase
transformations on the grassmanian coordinates of the 4D, $\CN=1$ superspace 
and would give $\CN=(0,1)$ CC sigma--models.
In the rest of the paper we concentrate only on the $(1,0)$ case. As noted in 
\cite{0508187}, the $(0,1)$ case can be recovered by simply 
doing the change $\pa\leftrightarrow-\pab$ wherever $\pa$ and $\pab$ appear.

Reduced in components the action (\ref{kahler6DCC}) reads
\bea
&&\int d^6x\Bigg\{~K_{a\Bar{b}}\,\Big[-{1\over 2}\,\pa^{\a\ad}
\Bar{A}^{\Bar{b}}\pa_{\a\ad}A^a+
\Bar{F}^{\Bar{b}}F^a
-{i\over 2}\(\Bar{\psi}^{\Bar{b}}_\ad\pa^{\a\ad}\psi_\a^a+\psi_\a^a
\pa^{\a\ad}\Bar{\psi}^{\Bar{b}}_\ad\)
\Big] \non\\
&&~~~~~~~~~+{1\over 2}\,K_{ab\Bar{c}}\,\Big[\,\Bar{F}^{\Bar{c}}\psi^{a\a}
\psi^b_\a
+i(\pa^{\a\ad}A^b)\psi^a_\a\Bar{\psi}^{\Bar{c}}_\ad\,\Big] \non\\
&&~~~~~~~~~+{1\over 2}\,K_{c\Bar{a}\Bar{b}}\,\Big[\,F^c\Bar{\psi}^{\Bar{a}\ad}
\Bar{\psi}^{\Bar{b}}_\ad
-i(\pa^{\a\ad}\Bar{A}^{\Bar{b}})\psi^c_\a\Bar{\psi}^{\Bar{a}}_\ad\,\Big] \non\\
&&~~~~~~~~~+Q_{a(b)}\psi^{b\a}\pa\,\psi^a_\a+
\Big(Q_{b(a)}-Q_{a(b)}\Big)F^a\pa A^b\,
+{1\over 2}\,Q_{a(bc)}\,(\pa A^a)\,\psi^{b\a}\psi^c_\a \non\\
&&~~~~~~~~~+\Bar{Q}_{\Bar{a}(\Bar{b})}\Bar{\psi}^{\Bar{b}\ad}\pab\,
\Bar{\psi}^{\Bar{a}}_\ad
+\Big(\Bar{Q}_{\Bar{b}(\Bar{a})}-\Bar{Q}_{\Bar{a}(\Bar{b})}\Big)
\Bar{F}^{\Bar{a}}\pab\,\Bar{A}^{\Bar{b}}
+{1\over 2}\,\Bar{Q}_{\Bar{a}\,(\Bar{b}\Bar{c})}(\pab\,\Bar{A}^{\Bar{a}})\,
\Bar{\psi}^{\Bar{b}\ad}
\Bar{\psi}^{\Bar{c}}_\ad \non\\
&&~~~~~~~~~+{1\over 4}\,K_{ab\Bar{a}\Bar{b}}\,\psi^{a\a}\psi_\a^b
\Bar{\psi}^{\Bar{a}\ad}\Bar{\psi}^{\Bar{b}}_\ad~
\Bigg\}~~~,~~~~~~
\label{Kahler6DCC2}
\eea
where we have defined the tensors
\beq
K_{a_1\cdots a_p\Bar{b}_1\cdots\Bar{b}_q}\equiv
{\pa^{p+q}K(A,\Bar{A})\over\pa A^{a_1}\cdots\pa A^{a_p}
\pa\Bar{A}^{\Bar{b}_1}\cdots\pa\Bar{A}^{\Bar{b}_q}}~~~,
\eeq
\beq
Q_{a(b_1\cdots b_r)}\equiv{\pa^r Q_a(A)\over\pa A^{b_1}\cdots\pa A^{b_r}}
~~~,~~~
\Bar{Q}_{\Bar{a}(\Bar{b}_1\cdots \Bar{b}_r)}\equiv
{\pa^r \Bar{Q}_{\Bar{a}}(\Bar{A})\over\pa
\Bar{A}^{\Bar{b}_1}\cdots\pa \Bar{A}^{\Bar{b}_r}}~~~.
\eeq
The equations of motion for the auxiliary $F$--fields are algebraic as
in the free case 
\bea
F^a&=&-K^{a\Bar{b}}\,\Bigg[\,{1\over
2}\,K_{cd\Bar{b}}\,\psi^{c\a}\psi^d_\a+
\Big(\Bar{Q}_{\Bar{c}(\Bar{b})}-\Bar{Q}_{\Bar{b}(\Bar{c})}\Big)\pab\,
\Bar{A}^{\Bar{c}}\,\Bigg]
~~~,~~~~~~\\
{\Bar{F}}^{\Bar{a}}&=&-K^{b\Bar{a}}\,
\Bigg[\,{1\over 2}\,K_{b\Bar{c}\Bar{d}}\,\Bar{\psi}^{\Bar{c}\ad}
\Bar{\psi}^{\Bar{d}}_\ad+
\Big(Q_{c(b)}-Q_{b(c)}\Big)\pa A^c\,\Bigg]~~~,~~~~~~
\eea
where $K^{a\Bar{b}}$ is the inverse of the K\"ahler metric 
$K_{a\Bar{b}}$,
$K_{a\Bar{c}}K^{b\Bar{c}}=\d_a^b$ and
$K_{c\Bar{a}}K^{c\Bar{b}}=\d_{\Bar{a}}^{\Bar{b}}$.
Inserting the previous relations in (\ref{Kahler6DCC2}) we find the
action for the physical component fields. 
We divide it into 
three pieces with zero, two and four fermionic fields, respectively
\beq
S_{0f}=
~\int d^6x\Bigg[-{1\over
2}\,K_{a\Bar{a}}\,\pa^{\a\ad}\Bar{A}^{\Bar{a}}\pa_{\a\ad}A^a
-K^{a\Bar{a}}\Big(Q_{b(a)}-Q_{a(b)}\Big)
\Big(\Bar{Q}_{\Bar{b}(\Bar{a})}-\Bar{Q}_{\Bar{a}(\Bar{b})}\Big)\pab\,
\Bar{A}^{\Bar{b}}\pa
A^b\,\Bigg]~~~,~~~
\label{0f}\\
\eeq
\bea
S_{2f}
&=&-{1\over 2}\int d^6x\Bigg[
K_{a\Bar{a}}\,\Bar{\psi}^{\Bar{a}}_\ad i\pa^{\a\ad}\psi_\a^a+
K_{ab\Bar{b}}\,\Big(i\pa^{\a\ad}A^a\Big)\Bar{\psi}^{\Bar{b}}_\ad\psi^b_\a
+\Big(Q_{b(a)}-Q_{a(b)}\Big)
\psi^{b\a}\pa\,\psi^a_\a \non\\
&&~~~~~~~~~~~~~~~
+K^{a\Bar{a}}K_{bc\Bar{a}}\Big(Q_{d(a)}-Q_{a(d)}\Big)
\Big(\pa A^d\Big)\psi^{b\a}\psi^c_\a 
\non\\
&&~~~~~~~~~~~~~~~
+\Big(Q_{b(ac)}-Q_{a(bc)}\Big)
\Big(\pa A^a\Big)\psi^{b\a}\psi^c_\a+~\{\,{\rm h.\,c.\,}\}~\Bigg]~~~,~~~
\label{2f}\\~~~\non\\
S_{4f}
&=&{1\over 4}\int d^6x\Bigg[\Big(K_{ab\Bar{a}\Bar{b}}\,-\,
K^{c\Bar{c}}K_{ab\Bar{c}}K_{c\Bar{a}\Bar{b}}\Big)\psi^{a\a}\psi_\a^b
\Bar{\psi}^{\Bar{a}\ad}\Bar{\psi}^{\Bar{b}}_\ad~\Bigg]~~~.
\label{4f}
\eea
In these actions the structures of the K\"ahler geometry as required by
manifest 4D, $\CN=1$ SUSY appear:
In (\ref{2f}, \ref{4f}), besides the metric, we recognize the connections
and the curvature tensor of the K\"ahler manifold
\bea
&\G^a_{~bc}=K^{a\Bar{d}}K_{bc\Bar{d}}~~~,~~~
\G^{\Bar{a}}_{~\Bar{b}\Bar{c}}=K^{d\Bar{a}}K_{d\Bar{b}\Bar{c}}~~~,\non\\
&{\cal R}_{a\Bar{b}c\Bar{d}}=
K_{ac\Bar{b}\Bar{d}}\,-\,K^{r\Bar{s}}K_{ac\Bar{s}}K_{r\Bar{b}\Bar{d}}~~~.
\eea
Extra constraints on the geometrical structures come from requiring that
after integration on the auxiliary $F$--fields, the resulting action is 6D
Lorentz invariant. In particular the actions
(\ref{0f}, \ref{2f}, \ref{4f}) must be separately Lorentz invariant.

\paragraph{Bosonic action}

We start imposing 6D Lorentz symmetry for the pure bosonic action 
(\ref{0f}). 
In order to have manifest, linearly realized 6D Lorentz invariance we 
should be able to write it as
\beq
-\int d^6x\Bigg[\,K_{a\Bar{a}}\,
\pa^\mu\Bar{A}^{\Bar{a}}\pa_\mu A^a
\,\Bigg]=
-{1\over 2}\int d^6x\Bigg[\,K_{a\Bar{a}}\Big(
\pa^{\a\ad}\Bar{A}^{\Bar{a}}\pa_{\a\ad}A^a+
\pab\,\Bar{A}^{\Bar{a}}\pa A^{a}+
\pa\Bar{A}^{\Bar{a}}\pab A^a\Big)
\Bigg]~~.~
\label{sigmaBoson6D}
\eeq
To compare the action (\ref{0f}) with (\ref{sigmaBoson6D}) we 
re-write (\ref{0f}) as
\bea
S_{0f}&=&
-{1\over 2}\int d^6x\Bigg[\,K_{a\Bar{a}}\,\pa^{\a\ad}
\Bar{A}^{\Bar{a}}\pa_{\a\ad}A^a+
\widetilde{K}_{a\Bar{a}}\(\,\pab\,\Bar{A}^{\Bar{a}}\pa A^a+
\pa\Bar{A}^{\Bar{a}}\pab A^a\) \non\\
&&
~~~~~~~~~~~~~~~
+\widetilde{K}_{a\Bar{a}}\(\,\pab\,\Bar{A}^{\Bar{a}}\pa A^a-
\pa\Bar{A}^{\Bar{a}}\pab A^a\)
\Bigg]~~~,~~~~~~
\label{0f2}
\eea
where we have defined
\beq
\widetilde{K}_{a\bar{a}}\,\equiv\,
\Big(\,Q_{b(a)}-Q_{a(b)}\,\Big)\,K^{b\Bar{b}}\,
\Big(\,\Bar{Q}_{\Bar{b}(\Bar{a})}-\Bar{Q}_{\Bar{a}(\Bar{b})}\,\Big)\,\equiv\,
-\O_{ab}\,K^{b\Bar{b}}\,\Ob_{\Bar{b}\Bar{a}}~~,
\eeq
and
\beq
\O_{ab}\,\equiv\, \Big(Q_{b(a)}-Q_{a(b)}\Big)~~~,~~~
\Ob_{\Bar{a}\Bar{b}}\,\equiv\, \Big(\Bar{Q}_{\Bar{b}(\Bar{a})}-
\Bar{Q}_{\Bar{a}(\Bar{b})}\Big)~~.
\label{OObHyper}
\eeq
Matching (\ref{0f2}) with (\ref{sigmaBoson6D}) requires 
\bea
K_{a\Bar{a}}&=&\widetilde{K}_{a\Bar{a}}\,=\,
-\O_{ab}\,K^{b\Bar{b}}\,\Ob_{\Bar{b}\Bar{a}}~~~.
\label{constrCC}
\eea
The second line of (\ref{0f2}) then becomes
\bea
&& -{1\over 2}\int d^6x
\Big(K_{a\Bar{a}}\,\pab\,\Bar{A}^{\Bar{a}}\pa A^a+
K_{ab}\,\pab A^b\pa A^a-K_{ab}\,\pa A^b\pab A^a
-K_{a\Bar{a}}\,\pa\Bar{A}^{\Bar{a}}\,\pab A^a\Big) \non\\
&& =
-{1\over 2}\int d^6x
\Big[\,\pab\(\,K_{a}\pa A^a\,\)-\pa\(\,K_{a}\pab A^a\,\)\Big]~~~,
\label{Hyper1}
\eea
and it is explicitly a total derivative in six dimensions.

An interesting observation regarding the
total derivative term is that if we were to work in 5D \cite{BagXio} 
the second line of (\ref{0f2}) would be 
identically zero, even without imposing 
$K_{a\Bar{a}}=\widetilde{K}_{a\Bar{a}}$ since $\pa$ = $\pab$ in five 
dimensions.

To summarize, in order  to have 6D Lorentz invariance of (\ref{0f}) we need
only require the constraint (\ref{constrCC}) for the K\"ahler metric.
Note that if we interpret $ Q_{a}$ as a holomorphic
1-form connection for a U(1) bundle then clearly the quantity $\O_{ab}$ is its 
exterior derivative (i.e. its field strength - U(1) curvature).

We now study the consequences of the constraint (\ref{constrCC}).
Taking the determinant of both sides of (\ref{constrCC}) we have an expression 
that relates the determinant of the K\"ahler metric to the exterior derivative
of the holomorphic one-form
\bea
\det{K}=\det{\O}\,\det{K^{-1}}\det{\Ob}~&
\Longrightarrow&
{\Big [} \,  \det{K} \, {\Big ]}^2 ~=~ \det{\O}\,\det{\Ob}=| \, \det{\O} \,|^2 
\, \, ~~~ \non\\
~\Longrightarrow~~~
 {\rm {Tr}} [ \ln(K) \,] &=& \fracm 12 \, {\Big [} \,  {\rm {Tr}}[ \, \ln(\O ) 
\, ] ~+~
{\rm {Tr}} [ \, \ln(\Ob) \, ]  \, {\Big ]} ~~~.
\label{squaredet}
\eea
These equations can be re-written in terms of the K\" ahler potential as
a nonlinear $d$-th order differential equation 
\bea
 && [ \det ( \pa_a \pa_{\Bar{b}}\, K) \,] ~=~  | \, \det{\O}\, \, |   
\label{Monge-Ampere}\\ 
 &&\fracm 1{d!}\,  \e^{a_1 \, a_2 \dots \, a_d } \,  \e^{{\Bar{b}{}_1 } \, 
{\Bar{b}{}_2 }
 \, \dots \, {\Bar{b}{}_d } }  \, ( \pa_{a_i}  \pa_{{\Bar{b}{}_i }}\, K) \,
 \cdots \, ( \pa_{a_i} \pa_{{\Bar{b}{}_i }}\, K) ~=~   | \, \det{\O}\, \, |  
  ~~~,
\label{DEQklr}
\eea
where $d$ is the number of the chiral doublets present in the action.
After the introduction of a new variable ${\cal K}(\Phi, \, {\Bar \Phi})$ via 
the equation 
$K \,=\, \Phi^a {\Bar \Phi}{}_{\bar a} \,+\, {\cal K}$ this leads to a 
nonlinear differential equation for $\cal K$,
\bea
  \det ( \, \d{}_{a \, {\bar b}} ~+~
 \pa_a \pa_{\Bar{b}}\, {\cal K})  ~=~ | \,  \det{\O}\, | ~~~.
\label{DEQklr2}
\eea

Since our manifold is K\"ahler we can express the Ricci tensor in terms of the
determinant of the metric as
$R_{a\Bar{b}}\,=\,\pa_a \pa_{\Bar{b}}\, [\,\ln(\det K)]$ and from 
(\ref{squaredet}) it follows
\beq
R_{a\Bar{b}}\,=\,
{1\over 2}\,\pa_a \pa_{\Bar{b}}\,[\,\ln{(\det{\O})} + \ln{(\det{\Ob})}]
\,=\,0~~~.
\eeq
Our manifold is then Ricci flat.

We note that, through a holomorphic change of coordinates, the 
$\det{\O}$ can be always chosen to be unimodular. Then the previous description
(\ref{squaredet})--(\ref{DEQklr2}) is equivalent to Monge-Amp\`ere 
equation $\det{(\pa_a \pa_{\Bar{b}}K)}=1$ which characterizes the Ricci 
flatness of our target space.

It is known that relations (\ref{OObHyper}, \ref{constrCC}) imply
moreover the stronger constraint on the target space geometry to be 
hyper-K\"ahler \cite{HuKaLiRo}. In fact, we introduce 
\bea
&\O^{ac}\,\O_{cb}\,=\,\d^a_{b}~~~,~~~
\Ob^{\Bar{a}\Bar{c}}\,\Ob_{\Bar{c}\Bar{b}}\,=\,\d^{\Bar{a}}_{\Bar{b}}~~~
\eea
and define
\bea
&&\O^{\Bar{a}}_{~b}\,\equiv \,K^{c\Bar{a}}\,\O_{cb}\,=\,
-\,\Ob^{\Bar{a}\Bar{c}}K_{b\Bar{c}}~~~,~~~~~~\\
&&\O^{a}_{~\Bar{b}}\,\equiv \,K^{a\Bar{c}}\,\Ob_{\Bar{c}\Bar{b}}\,=\,
-\,\O^{ac}K_{c\Bar{b}}~~~,~~~~~~
\eea
satisfying 
\beq
\O^{a}_{~\Bar{c}}\,\O^{\Bar{c}}_{~b}\,=\,-\,\d^{a}_{b}~~~,~~~
\O^{\Bar{a}}_{~c}\,\O^{c}_{~\Bar{b}}\,=\,-\,\d^{\Bar{a}}_{\Bar{b}}~~~.
\label{csHyper}
\eeq
It then follows
\bea
&\pa_a\,\O_{bc}\,=\,K_{ab\Bar{b}}\,\O^{\Bar{b}}_{~c}\,-\,
K_{ac\Bar{b}}\,\O^{\Bar{b}}_{~b}~~~\Longrightarrow~~~\Dc_a\,\O_{bc}\,=\,0~~~,\\
&\pa_{\Bar{a}}\,\O_{\Bar{b}\Bar{c}}\,=\,K_{b\Bar{a}\Bar{b}}\,
\O^{b}_{~\Bar{c}}\,-\,K_{b\Bar{a}\Bar{c}}\,\O^{b}_{~\Bar{b}}~~~
\Longrightarrow~~~\Dc_{\Bar{a}}\,\Ob_{\Bar{b}\Bar{c}}\,=\,0~~~,
\label{covConst}
\eea
and $\O_{ab}$, $\Ob_{\Bar{a}\Bar{b}}$, $\O^{a}_{~\Bar{b}}$ and 
$\O^{\Bar{a}}_{~b}$ are covariantly constant.
A triplet of covariantly constant complex structures
can be then introduced as in \cite{HuKaLiRo}--\cite{WitRocVan}
\beq
J^{1}=
\(\begin{array}{cc}0&\O^{a}_{~\Bar{b}}\\\O^{\Bar{a}}_{~b}&0\end{array}\)
~~~,~~~
J^{2}=
\(\begin{array}{cc}0&i\O^{a}_{~\Bar{b}}\\-i\O^{\Bar{a}}_{~b}&0\end{array}\)
~~~,~~~
J^{3}=
\(\begin{array}{cc}i\d^{a}_{~b}&0\\0 &-i\d^{\Bar{a}}_{~\Bar{b}}\end{array}\)
~~~.
\eeq
which define the 
quaternionic structure of an hyper-K\"ahler manifold
\beq
J^{\mu}\,J^{\nu}\,=\,-\,\d^{\mu\nu}\,+\,\epsilon^{\mu\nu\rho}\,J^{\rho}~~~,
\label{quat}
\eeq
Therefore, the request for the on--shell bosonic action to
be 6D Lorentz invariant implies the target space to be hyper-K\"ahler.

\paragraph{Fermionic actions} 

We now investigate the Lorentz invariance of the fermionic actions (\ref{2f},
\ref{4f}). As we are going to prove, the hyper-K\"ahler condition
for the target manifold
is sufficient to automatically provide 6D Lorentz invariance also for the  
fermionic actions, once properly defined the 6D, $(1,0)$ 
spinors\footnote{for our $(1,0)$ spinor conventions see \cite{0508187}.} 
as obtained from the 4D spinor components of the (anti)chiral superfields 
($\Bar{\Psi}^{\Bar{a}}$) $\Psi^a$. The correct choice of 6D spinors is the one 
suggested by the dimensional reduction of \cite{SieTow} and used also in the 
recent five dimensional analogue of our investigation \cite{BagXio}
\beq
\Psi^{a\tilde{\a}}=\(\begin{array}{c} \psi^{a\a}\vspace{1ex}\\
\O^{a}_{~\Bar{b}}\Bar{\psi}^{\Bar{b}\ad}\end{array}\)~~~,~~~
\Bar{\Psi}^{\Bar{a}\tilde{\a}}=
\(\begin{array}{c} -\O^{\Bar{a}}_{~b}\psi^{b\a}\vspace{1ex}\\
\Bar{\psi}^{\Bar{a}\ad}\end{array}\)=
-\O^{\Bar{a}}_{~b}\Psi^{b\tilde{\a}}~~~.
\label{defSpinorSM}
\eeq
Note that this is also the choice that gives a symplectic Majorana--Weyl 
structure to the 6D spinor. In fact, 
$(\Psi^{a\tilde{\a}})^*=\Bar{\Psi}^{\Bar{a}\dot{\tilde{\a}}}=
{\cal C}^{\dot{\tilde{\a}}}_{~~\tilde{\b}}
\,\O^{\Bar{a}}_{~~b}\Psi^{b\tilde{\b}}$ where 
${\cal C}^{\tilde{\a}}_{~~\dot{\tilde{\b}}}
{\cal C}^{\dot{\tilde{\b}}}_{~~\tilde{\g}}=-
\d^{\tilde{\a}}_{~~\tilde{\g}}$.\\
Now, using the following relations due to the hyper-K\"ahler structure
\bea
&{\cal R}_{a\Bar{a}b\Bar{b}}\,\O^{b}_{~\Bar{c}}\,=\,
\pa_a\Big(\Gamma^{\Bar{d}}_{~\Bar{a}\Bar{b}}\,\Ob_{\Bar{d}\Bar{c}}\Big)\,=\,
\pa_a\Big(\Gamma^{\Bar{d}}_{~\Bar{a}\Bar{c}}\,\Ob_{\Bar{d}\Bar{b}}\Big)\,=\,
{\cal R}_{a\Bar{a}b\Bar{c}}\,\O^{b}_{~\Bar{b}}~~~,\\
&{\cal R}_{a\Bar{a}b\Bar{b}}\,\O^{\Bar{b}}_{~c}\,=\,
\pa_{\Bar{a}}\Big(\Gamma^{d}_{~ab}\,\O_{dc}\Big)\,=\,
\pa_{\Bar{a}}\Big(\Gamma^{d}_{~ac}\,\O_{db}\Big)\,=\,
{\cal R}_{a\Bar{a}c\Bar{b}}\,\O^{\Bar{b}}_{~b}~~~,
\eea
we find that the two and four fermions actions (\ref{2f}, \ref{4f}) can be 
re-written as
\bea
S_{2f}&=&-{1\over 2}\int d^6x\,K_{a\Bar{a}}\Bigg[
~\Bar{\psi}^{\Bar{a}}_\ad \,i\pa^{\a\ad}\psi_\a^a\,+\,
\Gamma^a_{bc}\,\Big(i\pa^{\a\ad}A^c\Big)\,\Bar{\psi}^{\Bar{b}}_\ad\psi^b_\a\,
\non\\
&&~~~~~~~~~~~~~~~~~~~~
+\,\O^{\Bar{a}}_{~b}\,\psi^{b\a}\pa\psi^a_\a\,+\,
\Gamma^a_{cd}\Big(\pa A^d\Big)\,\O^{\Bar{a}}_{~b}\,\psi^{b\a}\psi^c_\a
~+~\{\,{\rm h.\,c.\,}\}\,\Bigg] \non\\
&=&{1\over 4}\int d^6x\,K_{a\Bar{a}}\Bigg[~
\Bar{\Psi}^{\Bar{a}\tilde{\a}}i\pa_{\tilde{\a}\tilde{\b}}\Psi^{a\tilde{\b}}
\,+\,
\Bar{\Psi}^{\Bar{a}\tilde{\a}}\Gamma^{a}_{bc}
\(i\pa_{\tilde{\a}\tilde{\b}}A^b\)\Psi^{c\tilde{\b}} 
\non\\
&&~~~~~~~~~~~~~~~~~~
+\,\Psi^{a\tilde{\b}}i\pa_{\tilde{\a}\tilde{\b}}\Bar{\Psi}^{\Bar{a}\tilde{\a}}
\,+\,\Psi^{a\tilde{\b}}\Gamma^{\Bar{a}}_{\Bar{b}\Bar{c}}
\(i\pa_{\tilde{\a}\tilde{\b}}\Bar{A}^{\Bar{b}}\)\Bar{\Psi}^{\Bar{c}\tilde{\a}}
\,\Bigg]~~~,
\label{2fFin}
\\~~~\non\\
S_{4f}&=&-{1\over 24}\int d^6x\
{\cal R}_{a\Bar{a}b\Bar{b}}\,\e_{\tilde{\a}\tilde{\b}\tilde{\g}\tilde{\d}}\,
\Psi^{a\tilde{\a}}\,\Psi^{b\tilde{\b}}\,
\Bar{\Psi}^{\Bar{a}\tilde{\g}}\,\Bar{\Psi}^{\Bar{b}\tilde{\d}}~~~.
\label{4fFin}
\eea
and Lorentz invariance become manifest. We have then found that 
6D Lorentz invariance requires the target space to be hyper-K\"ahler.
Under this condition, the sum of actions $(S_{0f}+S_{2f}+S_{4f})$ is also 
on--shell $\CN=(1,0)$ supersymmetric \cite{SieTow}.

Our sigma--model, being written in 4D $\CN=1$ superspace, has manifest 
4D supersymmetry. When the  hyper-K\"ahler conditions are satisfied, the action
is also on--shell invariant under the following transformations
\bea
\d_{\eta_2}\Psi^a=\Db^2\Big[\,\O^{ab}K_b\,
(\te^\a\eta_{2\a}+\teb^\ad\Bar{\eta}_{2\ad})\Big]
~~,~~
\d_{\eta_2}\Bar{\Psi}^{\Bar{a}}=
D^2\Big[\,\Ob^{\Bar{a}\Bar{b}}K_{\Bar{b}}\,
(\te^\a\eta_{2\a}+\teb^\ad\Bar{\eta}_{2\ad})
\Big]~.~~~
\label{SUSY2}
\eea
The 6D, $\CN=(1,0)$ algebra, once written in a 4D formalism, is 
equivalent to a 4D, $\CN=2$ SUSY algebra with a complex central charge 
\cite{0508187}. The transformations (\ref{SUSY2}) give exactly 
the second supersymmetry of the 4D, $\CN=2$ 
algebra. In fact, it can be seen that the commutator of two transformations
$[\d_{\eta_2},\d_{\zeta_2}]\Psi^a$ closes off--shell, and the
commutator of a transformation (\ref{SUSY2}) with a 4D $\CN=1$ 
transformation closes on--shell as 
$[\d_{\eta_2},\d_{\zeta_1}]\Psi^a=\pa\,\Psi^a
(\zeta_1^\a\eta_{2\a}+\Bar{\zeta}_1^\ad\Bar{\eta}_{2\ad})$ and 
$[\d_{\eta_2},\d_{\zeta_1}]\Bar{\Psi}^{\Bar{a}}=\pab\,\Bar{\Psi}^{\Bar{a}}
(\zeta_1^\a\eta_{2\a}+\Bar{\zeta}_1^\ad\Bar{\eta}_{2\ad})$ 
on the extra dimensions.
These properties are the natural extension to six dimensions of what happens
for five--dimensional CC sigma--models \cite{BagXio}.

\setcounter{equation}{0}
\section{6D sigma-models from projective superspace}

Up to now we have studied 6D supersymmetric sigma--models using a 
partially on--shell formalism which keeps 4D, $\CN=1$ SUSY manifest, 
being the target space coordinates described by 4D (anti)chiral superfield. 
This description is convenient due to the simplicity of the 4D, $\CN=1$ 
superspace structures but it has the disadvantage to realize only on--shell 
invariance under the whole 6D superpoincar\'e group. 

If we are interested in off--shell 6D superpoincar\'e invariant formulations, 
the most powerful description is harmonic superspace 
\cite{HarmonicSuperspace}--\cite{HarmonicSigma}
with eight supercharges which realize 6D, $\CN=1$ SUSY and SU$(2)$ 
automorphism group.  However, as we have emphasized previously, such 
constructions and approaches, at the quantum level, are necessarily bedeviled 
with harmonic divergences that make higher loop calculations ambiguous.  
Indeed, there presently does not exist a proof that such ambiguities can be 
removed to all orders of perturbation theory.

An alternative formulation which guarantees manifest off--shell supersymmetry 
for theories with eight supercharges can be obtained by using the projective 
superspace technique
\cite{ProjectiveSuperspace}--\cite{ProjectiveSuperspaceGK}. 
The two off--shell formulations are strictly related 
\cite{ProjectiveSuperspaceK} and the main difference is that the projective 
superspace approach has only a U$(1)$ subgroup linearly realized, out of the 
SU$(2)$ automorphism. The interesting property of projective 
superspace is that it naturally provides a reduction to 4D, $\CN=1$ superspace
which the harmonic approach does not admit.

Since in this paper we are interested in studying properties of 6D 
supersymmetric sigma--models with target space geometry parametrized by 4D, 
$\CN=1$ superfields, the projective superspace approach seems to be the most 
natural one.
A similar analysis has been recently performed for the 5D case in a series of 
papers \cite{projective5d}.

We start reviewing the definitions and properties of projective 
superspace in 6D \cite{GrunLind,0508187}. We focus on the reduction to 4D, 
$\CN=1$ superspace following the lines of our recent paper \cite{0508187}
(For conventions we refer the reader to this reference).

The algebra of the $\CN=(1,0)$ supercovariant derivatives is
\beq
\{D^{a\tilde{\a}},D^{b\tilde{\b}}\}=
\e^{ab}i\pa^{\tilde{\a}\tilde{\b}}
~~~,
\label{6dSUSYsupcovdev}
\eeq
where $\e^{ab}$ is the invariant tensor of the SU($2$) automorphism
group of the $\CN=(1,0)$ algebra and the derivatives
$D^{a\tilde{\a}}$ are $(1,0)$ Weyl spinors satisfying a
SU($2$)--Majorana condition \cite{SUSYKHSTK6d}. Now we extend the 6D
superspace parametrized by $Z=(x^\mu,\te_{a\tilde{\a}})$ with a projective
complex variable $\z\in{\mathbb{C}^*}$. In analogy with the 4D case we  
define the projective supercovariant derivatives as
\bea
\Dc^{\tilde{\a}}(\z)=u_a\Dc^{a\tilde{\a}}~~~,~~~
\D^{\tilde{\a}}(\z)=v_a\Dc^{a\tilde{\a}}~~~;~~~
u_a=(1,\z)~~~,~~~v_a=\(-{1\over\z},1\)~~~,
\label{projcovdev}
\eea
satisfying
\beq
\{\Dc^{\tilde{\a}},\Dc^{\tilde{\b}}\}=0~~~,~~~
\{\D^{\tilde{\a}},\D^{\tilde{\b}}\}=0~~~,~~~
\{\Dc^{\tilde{\a}},\D^{\tilde{\b}}\}=-2i\pa^{\tilde{\a}\tilde{\b}}~~~.
\eeq
We define superfields living in projective superspace as superfields 
holomorphic in $\z$ 
\beq
\Xi(Z,\z)=\sum_{n=-\infty}^{+\infty}\Xi_n(Z)\z^n~~~,
\label{compProjSup}
\eeq
and satisfying
\beq
\Dc^{\tilde{\a}}\,\Xi(Z,\z)=0~~~.
\label{projSuperfield}
\eeq
Following Ref. \cite{0508187} we want to make the structures of 4D
superfields manifest. In terms of 4D spinorial coordinates the 6D superspace is
parametrized by $Z=(x^\mu,\te^{a\a},\teb^{\ad}_a)$ and the 
algebra (\ref{6dSUSYsupcovdev}) is rewritten as 
\bea
\{D_{a\a},D_{b\b}\}=\e_{ab}C_{\a\b}\pab~~~,~~~
\{\Db^a_{\ad},\Db^b_{\bd}\}=\e^{ab}C_{\ad\bd}\pa~~~,~~~
\{D_{a\a},\Db^b_{\bd}\}=\d_a^bi\pa_{\a\bd}~~~.
\label{6dSUSYsupcovdev4Df}
\eea
It is interesting to note that this is equivalent to the algebra of 4D, 
$\CN=2$ SUSY with a complex central charge 
\cite{SUPERSPACE}. In 4D notations, the projective supercovariant derivatives
are
\bea
\Dc^{\tilde{\a}}=\(\begin{array}{c}\Dc^\a\vspace{1ex}\\\Dcb^\ad\end{array}\)=
\(\begin{array}{c}\z D_1^\a-D_2^\a\vspace{1ex}\\ 
\Db^{1\ad}+\z\Db^{2\ad}\end{array}\)~~~,~~~
\D^{\tilde{\a}}=\(\begin{array}{c}\D^\a\vspace{1ex}\\\Bar{\D}^\ad\end{array}\)=
\(\begin{array}{c}D_1^\a+{1\over \z}D_2^\a\vspace{1ex}\\ 
\Db^{2\ad}-{1\over \z}\Db^{1\ad}\end{array}\)~~~.~~
\label{projcovdevIN4D}
\eea
Then, from the definition (\ref{projSuperfield}), projective superfields 
satisfy
\bea
\Dc_\a(\z)\Xi=0=\Dcb_\ad(\z)\Xi~~~\Longleftrightarrow~~~ D_{2\a}\Xi=\z
D_{1\a}\Xi~~~,~~~\Db^1_\ad\Xi=-\z\Db^2_{\ad}\Xi~~~,
\label{constrProjSupf}
\eea
and the component superfields (\ref{compProjSup}) are constrained by
\bea
D_{2\a}\Xi_{n+1}=D_{1\a}\Xi_{n}~~~,~~~
\Db^2_{\ad}\Xi_{n}=-\Db^1_{\ad}\Xi_{n+1}~~~.
\label{compConstrProjSupf}
\eea
The above constraints fix the dependence of the $\Xi_n$ on half of the 
Grassmannian coordinates
of the superspace. The superfields $\Xi_n$ can then be considered as 
superfields living on a $\CN=1$ superspace with  
$\te^\a=\te^{1\a}$, $\teb^\ad={\Bar \q}_1^\ad$
\cite{ProjectiveSuperspace}--\cite{ProjectiveSuperspaceGK}, \cite{0508187} 
and we have a natural reduction of
6D, $\CN=(1,0)$ multiplets to 4D, $\CN=1$ superfields.\\
In projective superspace the natural conjugation 
operation combines complex conjugation with the antipodal map on the Riemann
sphere ($\z\to-1/\z$) and acts on projective superfields as
\beq
\Xis=\sum_{n=-\infty}^{+\infty}\Xis_n\,\z^n=
\sum_{n=-\infty}^{+\infty}(-1)^n\Xib_{-n}\,\z^n~~~.
\label{smileconjugation}
\eeq

Defining 
$\D^4={1\over 24}\e_{\tilde{\a}\tilde{\b}\tilde{\g}\tilde{\d}}\D^{\tilde{\a}}
\D^{\tilde{\b}}\D^{\tilde{\g}}\D^{\tilde{\d}}$, 
manifestly 6D $\CN=(1,0)$ SUSY invariant actions have the general 
form\footnote{We use the relations $\D^\a=2D^\a-{1\over\z}\Dc^\a$, 
$\Bar{\D}^\ad=-{2\over\z}\Db^\ad+{1\over\z}\Dcb^\ad$ which imply that
$\D^4=-16{1\over \z^2}D^2\Db^2$ \newline $~~~~~~$
when it acts on projective superfields and is integrated on the 6D 
space-time coordinates.}
\cite{GrunLind,0508187}
\beq
-\int d^6x\ggl\oint_C{\z d\z\over 32\pi i}\,\D^4\,
{\cal{L}}(\Xi,\Xis,\z)\Big|\ggr
=\int d^6xd^4\te\ggl\oint_C{d\z\over2\pi i\z}
\,{\cal{L}}(\Xi,\Xis,\z)\ggr~~~,
\eeq
where ${\cal{L}}(\Xi,\Xis,\z)$ is real under the
$\smile$-conjugation of (\ref{smileconjugation}) and $C$ is a contour
around the origin of the complex $\z$--plane.

The general classification of multiplets in projective superspace is 
based on the analyticity properties of the projective superfields in the 
$\z$--plane \cite{ProjectiveSuperspace}--\cite{ProjectiveSuperspaceK} 
and it is essentially not affected by the dimensions of the space--time. 
What different dimensions affect is the original SUSY algebra with
eight supercharges which are used to define the projective superspace. 
Note that the 6D case is interesting in this regard, six being the largest 
dimension in which hypermultiplets with only $(0, \frac12)$ degrees of 
freedom can be defined. Therefore, it can be considered as the parent 
(up to issues involving 'twists' and such dualities) of all lower dimensional 
theories with only $(0, \frac12)$ multiplets constructed by dimensional 
reduction.

Now, we consider a particular class of examples built using the 6D 
polar multiplet \cite{0508187} defined by (ant)artic superfields 
focusing on the reduction from projective superfields to
4D, $\CN=1$ superfields degrees of freedom.
It is an interesting feature of the 4D and 5D
projective superfields to provide coordinates for natural extensions of rigid 
$\CN=1$ K\"ahler nonlinear sigma--models to the $\CN=2$ 
cases \cite{ProjectiveSuperspaceGK,projective5d}. Adapting
these extensions to the 6D projective superspace it is
straightforward to find the same geometrical structures. 

We start by considering a 4D $\CN=1$ rigid supersymmetric
sigma--model \cite{zumino}
\beq
\int d^4xd^4\te
~K(\F^I,\Fib^{\Bar{I}})~~~,
\label{N14dsm}
\eeq
with $K$ the
K\"ahler potential of the target space K\"ahler manifold ${\cal
M}$ parametrized by the scalar components of $\F^I$
($\Fib^{\Bar{I}}$). In analogy to the 4D case we define a 6D
$\CN=(1,0)$ sigma--model on ${\cal M}$
\beq
\int
d^6xd^4\te\ggl\oint_C{d\z\over 2\pi i \z}
K(\Y^I(\z),\Ys^{\Bar{I}}(\z))\ggr~~~.
\label{N(1,0)6dsm}
\eeq
where $K$ is a function of the 6D (ant)artic projective 
superfields ($\Ys^{\Bar{I}}$) $\Y^I$ defined by the following power series
\bea
\Y^I=\sum_{n=0}^{+\infty}\Y^I_n\,\z^n&,&
\Ys^{\Bar{I}}=\sum_{n=0}^{+\infty}(-1)^n\Yb^{\Bar{I}}_n\,{1\over \z^n}~~~.
\label{(ant)artic}
\eea
The action (\ref{N(1,0)6dsm}) is invariant under the global
U$(1)$ transformation
\beq
\Y(\z)~~~\to~~~\Y(e^{i\a}\z)~~~~~~\Longleftrightarrow~~~~~~
\Y_n~~~\to~~~e^{in\a}\Y_n ~~~.
\label{globalU1}
\eeq
Due to the truncation of the series, the
$\CN=1$ constraints on the component superfields $\Y^I_n$, $\Yb^{\Bar{I}}_n$ 
are
\beq
\Db_\ad\Y^I_0=0~~~,~~~\Db^2\Y^I_1=\pa\,\Y^I_0~~~~~~;~~~~~~
D_\a\Yb^{\Bar{I}}_0=0~~~,~~~D^2\Yb^{\Bar{I}}_1=\pab\,\Yb^{\Bar{I}}_0~~~,
\label{polarConstr}
\eeq
with $\Y^I_n$, $\Yb^{\Bar{I}}_n$ ($n>1$) unconstrained $\CN=1$ superfields. 
The constraints
(\ref{polarConstr}) define a set of 6D chiral--nonminimal (CNM) hypermultiplets
\cite{0508187}
given by $\Y^I_0=\F^I$ and $\Y^I_1=\S^I$ extended 
with an infinite number of auxiliary superfields.

We observe that the action (\ref{N(1,0)6dsm})
has the same properties of the 4D, $\CN=1$ case (\ref{N14dsm}). It is invariant
under K\"ahler transformations
\bea
K(\Y,\Ys)&\longrightarrow&K(\Y,\Ys)+\L(\Y)+\Bar\L(\Ys)~~~,
\eea 
and holomorphic reparametrizations of the K\"ahler manifold
$\Y^I~~~\longrightarrow~~~f^I(\Y^J)$.

The physical superfields
\bea ~~\Y^I(\z)\Big|_{\z=0} ~=~
\F^I~~~~~~&,&~~~~~~ {{\rm d}\Y^I(\z)\over {\rm d}\z} \Big|_{\z=0}
~=~ \S^I ~~~,
\eea
of the 6D CNM hypermultiplet can be
regarded as parameters of the tangent bundle $T{\cal M}$ of the
K\"ahler manifold ${\cal M}$.

The simplest example concerns a flat one--dimensional manifold 
with $K=\Fib\F$ in (\ref{N14dsm}). In this case the action
(\ref{N(1,0)6dsm}) becomes
\beq
\int
d^6xd^4\te\ggl\oint_C{d\z\over 2\pi i\z}\,\Ys\Y\ggr= \int
d^6xd^4\te\ggl\Fib\F-\Sb\S+\sum_{n=2}^{+\infty}(-1)^n\Yb_n\Y_n\ggr~~~.
\label{polarAction}
\eeq
After integrating out the auxiliary superfields $\Y_n$, $\Yb_n$ with $n>1$,
we have $\int d^6xd^4\te[\Fib\F-\Sb\S]$ which is the action for a free 6D
$\CN=(1,0)$ CNM hypermultiplet which has been investigated in \cite{0508187}.
In particular, it is dual 
to the free CC formulation (\ref{S0CC}). 

The analysis of the free system can be extended to the non--trivial cases
(\ref{N(1,0)6dsm}). 
We need eliminate the auxiliary superfields of the polar
hypermultiplet.
This can be done exactly as in the 4D 
case \cite{ProjectiveSuperspaceGK} where we refer the reader for 
details (see also \cite{0602277} for recent applications).
The action we are left with has the following form
\bea
&&S_{CNM}(\F^I,\Fib^{\Bar{I}},\S^I,\Sb^{\Bar{I}}) = \int
d^6xd^4\te\Bigg\{ K(\F,\Fib)-g_{I\Bar{J}}(\F,\Fib)\S^I\Sb^{\Bar{J}}
\non\\
&&
~~~~~~+\sum_{p=2}^{+\infty}{\cal R}_{I_1 \cdots I_p {\Bar
J}_1 \cdots {\Bar J}_p }(\F,\Fib)\S^{I_1}\cdots\S^{I_p}\Sb^{{\Bar
J}_1 } \cdots\Sb^{{\Bar J}_p}\Bigg\}~~~,
\label{CNMsm}
\eea
where the
tensors ${\cal R}_{I_1 \cdots I_p {\Bar J}_1 \cdots {\Bar J}_p }$
are functions of the Riemann curvature $R_{I\Bar{J}K\Bar{L}}$ and
its covariant derivatives. All the terms contain equal powers of 
$\S$ and $\Sb$ as a consequence of the invariance under (\ref{globalU1}).
It is worth the mention, that presently, there is in general not known a 
closed-form analytic expression for 
${\cal R}_{I_1 \cdots I_p {\Bar J}_1 \cdots {\Bar 
J}_p }(\F,\Fib)$.  A solution to this problem would represent a major
advance in understanding this class of problems.

The action (\ref{CNMsm}) describes a class of non--trivial 6D CNM sigma--models
which are guaranteed to be on--shell $\CN=(1,0)$ supersymmetric and
6D Lorentz invariant by construction. 

So far we have restricted our attention to the polar multiplet as
an extension of the 4D chiral multiplet. In particular, we have constructed 
6D, $\CN=(1,0)$ supersymmetric sigma--models defined over the tangent bundle 
$T{\cal M}$ of a K\"ahler manifold ${\cal M}$. In the four
dimensional case, using the projective superspace, in
\cite{ProjectiveSuperspaceGK,projective5d} an extension of the
rigid $c$--map \cite{CeFeGi} was proposed which allows us to obtain a
4D, $\CN=2$ hyper-K\"ahler manifold starting from a 4D special
K\"ahler geometry. The construction makes use of $O(2n)$
multiplets in projective superspace. Without giving any detail,
we note that, as follows from our previous discussion, the construction 
of $O(2n)$ 6D, $\CN=(1,0)$
hyper-K\"ahler sigma models along the lines of
\cite{ProjectiveSuperspaceGK} should work straightforwardly since
the dimensions of the space--time should not affect the superspace
structures which allow for that construction.

We conclude by noting that our previous analysis covers only a small set
of projective superspace sigma--models. The relevant property of actions of
the form (\ref{N(1,0)6dsm}) is that 
the auxiliary superfields integration procedure is quite well understood and
solved exactly for some non--trivial examples 
\cite{ProjectiveSuperspaceGK,0602277}. It is believed that all the 
hyper-K\"ahler metrics can be derived from
the most general polar multiplet action $K(\Y,\Ys,\z)$ 
with a non--trivial dependence on $\z$ \footnote{We thank Martin 
Ro\v cek for electronic correspondence on this point and for informing us
on a\\$~~~~~$ forthcoming proof of this claim \cite{LiRoToAppear}. In harmonic 
superspace it is known that all hyper-K\"ahler\\$~~~~~$
metrics can be found from the most general $q^+$ hypermultiplet action
\cite{HarmonicSuperspace,HarmonicSigma}.}. 
We expect that the CNM's would arise naturally also in the general projective 
superspace case and the 6D structure would be the same as in our present 
analysis.

\setcounter{equation}{0}
\section{6D, $\CN=(1,0)$ CNM sigma-models}

Six--dimensional projective superspace provides a powerful method to build a
class of 6D, $\CN=(1,0)$ supersymmetric nonlinear sigma--models whose partially
on--shell description is given in terms of CNM 4D, $\CN=1$ superfields. 
The projective 
superspace construction insures that the resulting CNM sigma--model is
on--shell 6D, $\CN=(1,0)$ supersymmetric and we expect the structure of 
the CNM target space geometry to arise naturally. However, the action 
(\ref{CNMsm}) for a 6D sigma--model as coming from projective superspace is not
the most general action consistent with the symmetries of the problem.  

In this section we investigate the most general class of CNM sigma--models we
can construct directly in
terms of 4D superfields and figure out the associated target space geometry,
as done in  section 2 for the CC case. 
In particular, we study how the defining tensors of the model 
are constrained by the demand of on--shell 6D, $\CN=(1,0)$ SUSY. 

Generalizing the free $\CN=(1,0)$ CNM action \cite{0508187}, we consider the 
following ansatz for the most general $(1,0)$ CNM sigma--model 
action, off--shell invariant under 4D SUSY and the 
Sl$(2,\mathbb{C})\times$U$(1)$ subgroup of the 6D Lorentz group
\beq
S = \int d^6x\Bigg[\int d^4\te\, G\Big(\F^a,\Fib^{\Bar{a}},\Sigma^k,
\Bar{\Sigma}^{\Bar{k}}\Big)+
\int d^2\te\, P_a\Big(\F^b\Big)\pa\F^a+
\int d^2\teb\,\Bar{P}_{\bar{a}}\Big(\Fib^{\Bar{b}}\Big)\pab\Fib^{\Bar{a}}
\,\Bigg]~~~,
\label{SCNMg}
\eeq
where the 
superfields $\F^a,\,\Fib^{\Bar{a}},\,\Sigma^k,\,
\Bar{\Sigma}^{\Bar{k}}$ are CNM satisfying
\bea
\Db_\ad\F^a=0~~~&,&~~~\Db^2\S^k=S^k_{a}(\F)\,\pa\,\F^{a}~~~,
\non\\
D_\a\Fib^{\Bar{a}}=0~~~&,&~~~D^2\Sb^k=
\Bar{S}^{\Bar{k}}_{\Bar{a}}(\Fib)\,\pab\,\Fib^{\Bar{a}}~~~.
\label{constrCNMg2}
\eea
The CNM models emerging from projective superspace correspond to the 
particular choice $P_a = 0$, $S=1$ and $G$ constrained to have
the form (\ref{CNMsm}).  

In trying to keep the discussion very general we allow the number of chiral 
($n_c$) and nonminimal ($n_{nm}$) superfields to be different, 
we generalize the nonminimal constraint by the introduction of the
tensor $S^a_k(\F)$ and add holomorphic
terms admitted by the symmetries of the theory.

Actually, the introduction of a holomorphic term is necessary whenever 
$n_{nm} < n_c$. As a particular example we mention the case of one free CC plus
one free CNM pairs ($n_{nm}=1$, $n_c=3$)
\beq
S = \int d^6x\Bigg[\int d^4\te\, [\Fib_+\F_++\Fib_-\F_-+\Fib\F-\Sb\S\,]
+\int d^2\te \, \F_+\pa\F_- + \int d^2\teb \, \Fib_+\pab\Fib_-\,\Bigg]~,
\eeq
\bea
\Db_\ad\F_\pm=\Db_\ad\F =0, ~~\,,~~
D_\a\Fib_\pm=D_\a\Fib=0~~\,,~~
\Db^2\S=\pa\,\F\,~~~,~~~D^2\Sb=\pab\,\Fib~~~.~~
\eea
While the completion to 6D of the CNM $(\F,\S)$ kinetic terms is provided by 
the non--trivial constraint  $\Db^2\S=\pa\F$, the completion of the kinetic 
terms for $\F_{\pm}$ makes use of the holomorphic term, as discussed in 
\cite{0508187} and in section 2.

Now, we go back to the general case (\ref{SCNMg}, \ref{constrCNMg2}).
In the CC case of section 2 we have first imposed the restoration of 6D 
Lorentz invariance on the bosonic 
part of the action with the auxiliary fields set on--shell.
The requirement of 6D Lorentz invariance constrains the target space to be 
hyper-K\"ahler and
this is sufficient to guarantee the on--shell invariance of the whole
action plus 6D, $\CN=(1,0)$ supersymmetry. We now follow the same approach
to constrain the tensors $G,\,P,\,S,\,\Bar{P}$ and
$\Bar{S}$ of the CNM sigma--models (\ref{SCNMg}, \ref{constrCNMg2}).

Having defined the component fields as in (\ref{componentsCNM}), we
reduce the action (\ref{SCNMg}) in components. 
The resulting action is much more complicated than the CC one and we refer the 
reader to  appendix B for the whole
component lagrangian (see eq. (\ref{ScomponentsCNM6D})). 

Before performing the auxiliary fields integration, it is useful to write the 
bosonic part of (\ref{ScomponentsCNM6D}) in a compact form introducing 
a vectorial/matricial notation. We define the following matrices ($G_a \equiv 
\frac{\partial G}{\partial \F^a}$)
\bea
M\equiv\(\begin{array}{cc} 0&G_{a\Bar{b}}\\
G_{\Bar{a}b}&0\end{array}\)~~~,~~~
N\equiv\(\begin{array}{cc} G_{ar}&G_{a\Bar{r}}\\
G_{\Bar{a}r}&G_{\Bar{a}\Bar{r}}\end{array}\)~~~,~~~
H\equiv\(\begin{array}{cc} G_{kr}&G_{k\Bar{r}}\\
G_{\Bar{k}r}&G_{\Bar{k}\Bar{r}}\end{array}\)~~~,~
\label{defMatr2}
\\~~~\non\\
{\cal S}\equiv\(\begin{array}{cc} S^k_{b}&0\\
0&\Bar{S}^{\Bar{k}}_{\Bar{b}}\end{array}\)~~~,~~~
P\equiv\(\begin{array}{cc} 1&0\\0&-1\end{array}\)~~~,~~~
P_\pm\equiv{1\over 2}(1\pm P)~~~,~~~~~~~~~~~
\label{defMatr3}
\\~~~\non\\
{\cal O}\equiv\(\begin{array}{cc} (G_kS^k_b+P_b)_{(a)}-(G_kS^k_a+P_a)_{(b)}
&-S^k_aG_{k\Bar{b}}\\
-\Bar{S}^{\Bar{k}}_{\Bar{a}}G_{\Bar{k}b}&
(G_{\Bar{k}}\Bar{S}^{\Bar{k}}_{\Bar{b}}+\Bar{P}_{\Bar{b}})_{(\Bar{a})}-
(G_{\Bar{k}}\Bar{S}^{\Bar{k}}_{\Bar{a}}+\Bar{P}_{\Bar{a}})_{(\Bar{b})}
\end{array}\)~~~.~~
\eea
in terms of which the bosonic component lagrangian becomes
\bea
S_{0f} &=& \int d^6 x \Bigg[-\,{1\over 4}\,\pa^{\a\ad}{\cal A}^T\,M\,
\pa_{\a\ad}{\cal A}\,+\,
{1\over 8}\,\pa^{\a\ad}{\cal B}^T\,[3H+PHP]\,\pa_{\a\ad}{\cal B} \non\\
&&~~~~~~~~~\,+\,{1\over 8}\,\pa^{\a\ad}{\cal A}^T\,[P,[P,N]]\,
\pa_{\a\ad}{\cal B}\,+\,
\pa{\cal A}^T\,P_+{\cal S}^TH{\cal S}P_-\,\pab{\cal A} \non\\
&&~~~~~~~~~\,+\,{1\over 2}\,{\cal F}^T\,M\,{\cal F}\,+\,
{1\over 8}\,{\cal H}^T\,[P,[P,H]]\,{\cal H}\,+\,
{1\over 4}\,{\cal F}^T\,[P,[P,N]]\,{\cal H} \non\\
&&~~~~~~~~~\,+\,{\cal F}^T\,P_+{\cal O}\,\pa{\cal A}\,+\,
\,{\cal F}^T\,P_-{\cal O}\,\pab{\cal A}\,+\,
\,{\cal H}^T\,P_+H{\cal S}P_+\,\pa{\cal A} \non\\
&&~~~~~~~~~\,+\,{\cal H}^T\,P_-H{\cal S}P_-\,\pab{\cal A}\,-\,
{\cal F}^T\,P_+{\cal S}^TH\,\pa{\cal B}\,-\,
\,{\cal F}^T\,P_-{\cal S}^TH\,\pab{\cal B} \non\\
&&~~~~~~~~~\,-\,{1\over 2}\,{\cal P}^T_{\a\ad}\,H\,{\cal P}^{\a\ad}\,+\,
{1\over 2}\,{\cal P}^T_{\a\ad}\,[P,N^T]\,i\pa^{\a\ad}{\cal A}\,+\,
{1\over 2}\,{\cal P}^T_{\a\ad}\,\{P,H\}\,i\pa^{\a\ad}{\cal B} \Big] ~~~.
\non\\
&&~~~~~~~
\label{boseCNM}
\eea
As usual, the equations of motion for the auxiliary fields are algebraic. 
Defining the matrices
\bea
&\widetilde{P}\equiv\(\begin{array}{c|c}P&0\\\hline0&P\end{array}\)~~~,~~~
{\cal G}\equiv\(\begin{array}{c|c}M&N\\\hline N^T&H\end{array}\)~~~,~~\\
&{\cal Z}=\({1\over 4}[\widetilde{P},[\widetilde{P},{\cal G}]]\)^{-1}~~~,~~~
{\cal X}_\pm={1\over 2}(1\pm\widetilde{P})
\(\begin{array}{c|c}{\cal O}&-{\cal S}^TH\\
\hline H{\cal S}P_\pm&0\end{array}\)~~~,~~~~~~
\eea
the solution to the equations of motion for the auxiliary fields read
\bea
\(\begin{array}{c}{\cal F}\\{\cal H}\end{array}\)&=&
-\,{\cal Z}{\cal X}_+\,\pa\(\begin{array}{c}{\cal A}\\{\cal B}\end{array}\)
\,-\,
{\cal Z}{\cal X}_-\,\pab\(\begin{array}{c}{\cal A}\\{\cal B}\end{array}\)~~~,
\non\\~~~\non\\
{\cal P}_{\a\ad}\,~&=&{1\over 2}\,H^{-1}[P,N^T]\,i\pa_{\a\ad}{\cal A}\,+\,
{1\over 2}\,H^{-1}\{P,H\}\,i\pa_{\a\ad}{\cal B}~~~.
\eea
Inserting back into (\ref{boseCNM}) and defining 
\bea
{\cal C}\equiv\(\begin{array}{c}{\cal A}\\{\cal B}\end{array}\)~~~,~~~
{\cal Y}\equiv{\cal X}_+^T{\cal Z}{\cal X}_-\,-\,
\(\begin{array}{c|c}P_+{\cal S}^TH{\cal S}P_-&0\\\hline
0&0\end{array}\) \label{Y}
~~~,~~~
\\~~~\non\\
{\cal K}\equiv
\(\begin{array}{c|c}
{\cal K}_1&{\cal K}_2\\\hline
{\cal K}_2^T&{\cal K}_3\end{array}\)
~~~,~~~
{\cal K}_1=M+{1\over 2}[P,N]H^{-1}[N^T,P]~~~,~~~~~~~~~~~~\label{calK}\\
{\cal K}_2={1\over 2}[N,P]H^{-1}\{P,H\}-{1\over 4}[P,[P,N]]
~~~,~~~
{\cal K}_3={1\over 2}HPH^{-1}PH-{1\over 2}H~~~,\label{calK1}
\eea
we find the following action for the bosonic physical fields 
\beq
S = -\,\int d^6x\Bigg[\,{1\over 4}\,\pa^{\a\ad}{\cal C}^T\,
{\cal K}\,\pa_{\a\ad}{\cal C}\,+\,
\pa{\cal C}^T\,{\cal Y}\,\pab{\cal C}\,\Bigg]~~~.~~~~
\label{0fCNM}
\eeq
The matrix ${\cal Y}$ defined in (\ref{Y}) is not symmetric. In order to 
proceed we need symmetrize it.
To this porpose we rewrite (\ref{0fCNM}) as
\bea
&&-~{1\over 4}\int d^6x\Bigg[\,
\pa^{\a\ad}{\cal C}^T~{\cal K}~\pa_{\a\ad}{\cal C}~+~
\pa{\cal C}^T~
\widetilde{{\cal K}}~\pab{\cal C}~+~
\pab{\cal C}^T~
\widetilde{{\cal K}}~\pa{\cal C}
\,\Bigg] \non\\
&&+~{1\over 2}\int d^6x\,\pa{\cal C}^T({\cal Y}-{\cal Y}^T)\,\pab{\cal C}~~~,
\label{0f2CNM}
\eea
where we have defined
\beq
\widetilde{{\cal K}} \equiv {\cal Y}+{\cal Y}^T 
~=~\widetilde{{\cal X}}^T{\cal Z}\widetilde{{\cal X}}\,-\,
\(\begin{array}{c|c}{1\over 4}[P,[P,{\cal S}^TH{\cal S}]]&0\\\hline
0&0\end{array}\)~~~,
\label{tildeK}
\eeq
and 
\beq
\widetilde{{\cal X}}~=~{\cal X}_+\,+\,{\cal X}_-\,=\,
\(\begin{array}{c|c}
{\cal O}&-{\cal S}^TH\\\hline
P_+H{\cal S}P_+\,+\,P_-H{\cal S}P_-&0
\end{array}\)~~~.
\eeq
Note that the structure of (\ref{0f2CNM}) is similar to the one 
for the CC case (see eq. (\ref{0f2})).
Therefore, by imposing the restoration of 6D Lorentz 
invariance we obtain the following condition
\bea
{\cal K}&=&\widetilde{{\cal K}}~~~.
\label{constr6DLorCNM}
\eea
As in the CC case we expect the 
constraint (\ref{constr6DLorCNM}) to be sufficient to make the second line of
(\ref{0f2CNM}) a total derivative and, more importantly,  
to provide on--shell 6D Lorentz invariance and 6D, $\CN=(1,0)$ SUSY
of the whole action. Unfortunately, in this case the direct proof is not
straightforward and we have not pursued the calculations up to the very end.

The constraint (\ref{constr6DLorCNM}), once written for each component of the 
two matrices,
gives rise to a system of equations which is much more intricated than
(\ref{constrCC}) for the CC case. Up to now we have not been able to solve it 
in general. We are going to provide the explicit solution only in the following
example.

\paragraph{Example: 4D target space.}

We consider the CNM sigma model describing the dynamics of one chiral and one 
nonminimal superfields defined by the action\footnote{We consider the simplest 
CNM constraint with $S(\F)=1$ since, in the present case, we \newline 
$~~~~~~$ can always eliminate the 
function $S(\F)$ by a redefinition of the nonminimal superfield,  \newline
$~~~~~~$ 
$\S\equiv S(\F)\S'$, which implies $\Db^2\S'=\pa\F.$}
\beq
S = \int d^6xd^4\te\,G(\F,\Fib,\S,\Sb)~~~,~~~
\Db^2\S\,=\,\pa\,\F~~~,~~~D^2\Sb\,=\,\pab\,\Fib~~~.
\label{Target4DCNM}
\eeq
In this case we can write explicitly all the quantities which enter our
equations (\ref{constr6DLorCNM}). In particular,  
${\cal K}$ (\ref{calK}, \ref{calK1}) has component matrices given by
\bea
{\cal K}_1~=~
\begin{small}
{1\over\det{H}}
\(\begin{array}{cc}
2G^2_{\F\Sb}G_{\S\S}&
G_{\F\Fib}(\det{H})+2G_{\F\Sb}G_{\S\Fib}G_{\S\Sb}\\
G_{\F\Fib}(\det{H})+2G_{\F\Sb}G_{\S\Fib}G_{\S\Sb}&
2G^2_{\S\Fib}G_{\Sb\Sb}
\end{array}\)
\end{small}
~~~,~~~
\\~~~\non\\
{\cal K}_2~=~
\begin{small}
{1\over\det{H}}
\(\begin{array}{cc}
2G_{\F\Sb}G_{\S\S}G_{\S\Sb}&
G_{\F\Sb}(G_{\S\S}G_{\Sb\Sb}+G^2_{\S\Sb})\\
G_{\S\Fib}(G_{\S\S}G_{\Sb\Sb}+G^2_{\S\Sb})&
2G_{\S\Fib}G_{\S\Sb}G_{\Sb\Sb}
\end{array}\)
\end{small}
~~~,~~~
~~~~~~~~~~~~~~\,~~\\~~~\non\\
{\cal K}_3~=~
\begin{small}
{1\over\det{H}}
\(\begin{array}{cc}
2G_{\S\S}G^2_{\S\Sb}&
G_{\S\Sb}(G_{\S\S}G_{\Sb\Sb}+G^2_{\S\Sb})\\
G_{\S\Sb}(G_{\S\S}G_{\Sb\Sb}+G^2_{\S\Sb})&
2G^2_{\S\Sb}G_{\Sb\Sb}
\end{array}\)
\end{small}
~~~,~~~~~~~~~~~~~~\,~~~~~
\eea
where $(\det{H})=(G_{\S\S}G_{\Sb\Sb}-G^2_{\S\Sb})$. Furthermore, we have
\bea
{\cal Z}\,=\,
\begin{small}
{1\over G_{\F\Fib}G_{\S\Sb}-G_{\F\Sb}G_{\S\Fib}}
\(\begin{array}{cc|cc}
0&G_{\S\Sb}&0&-G_{\S\Fib}\\
G_{\S\Sb}&0&-G_{\F\Sb}&0\\\hline
0&-G_{\F\Sb}&0&G_{\F\Fib}\\
-G_{\S\Fib}&0&G_{\F\Fib}&0
\end{array}\)
\end{small}~~~,
\eea
and 
$\widetilde{{\cal K}}=\(\begin{array}{c|c}
\widetilde{{\cal K}}_1&\widetilde{{\cal K}}_2\\\hline
\widetilde{{\cal K}}_2^T&\widetilde{{\cal K}}_3\end{array}\)$ 
in (\ref{tildeK}) becomes 
( $\tilde{k}\equiv (G_{\F\Fib}G_{\S\Sb}-G_{\F\Sb}G_{\S\Fib})$)
\bea
\widetilde{{\cal K}}_1&=&
{1\over \tilde{k}}
\begin{small}
\(\begin{array}{cc}
2G^2_{\F\Sb}G_{\S\S}&
G_{\F\Fib}G_{\S\S}G_{\Sb\Sb}+G_{\F\Sb}G_{\S\Fib}G_{\S\Sb}\\
G_{\F\Fib}G_{\S\S}G_{\Sb\Sb}+G_{\F\Sb}G_{\S\Fib}G_{\S\Sb}&
2G^2_{\S\Fib}G_{\Sb\Sb}
\end{array}\)~~~,~~~~~~~~
\end{small}\\~~~\non\\
\widetilde{{\cal K}}_2&=&
{1\over \tilde{k}}
\begin{small}
\(\begin{array}{cc}
2G_{\F\Sb}G_{\S\S}G_{\S\Sb}&
G_{\F\Sb}(G_{\S\S}G_{\Sb\Sb}+G^2_{\S\Sb})\\
G_{\S\Fib}(G_{\S\S}G_{\Sb\Sb}+G^2_{\S\Sb})&
2G_{\S\Fib}G_{\S\Sb}G_{\Sb\Sb}
\end{array}\)~~~,~~~
\end{small}\\~~~\non\\
\widetilde{{\cal K}}_3&=&
{1\over \tilde{k}}
\begin{small}
\(\begin{array}{cc}
2G_{\S\S}G^2_{\S\Sb}&
G_{\S\Sb}(G_{\S\S}G_{\Sb\Sb}+G^2_{\S\Sb})\\
G_{\S\Sb}(G_{\S\S}G_{\Sb\Sb}+G^2_{\S\Sb})&
2G^2_{\S\Sb}G_{\Sb\Sb}
\end{array}\)~~~,~~~
\end{small}
\eea
Now, imposing ${\cal K}=\widetilde{{\cal K}}$ as in (\ref{constr6DLorCNM}) the
only non--trivial condition we obtain is
\bea
G_{\S\S}G_{\Sb\Sb}-G^2_{\S\Sb}&=&G_{\F\Fib}G_{\S\Sb}-G_{\F\Sb}G_{\S\Fib}~~~.
\label{constrTarget4DCNM}
\eea
In this case one can check that this condition is sufficient for the second 
line of (\ref{0f2CNM}) to be a total derivative.
As we are going to show at the end of section 6 the condition 
(\ref{constrTarget4DCNM}) implies that the 4D target space geometry is 
hyper-K\"ahler. 

It is interesting to note that (\ref{constrTarget4DCNM}) 
is exactly the same constraint which was found in \cite{9803230} from the 
condition of vanishing one--loop beta--function for a 2D CNM sigma--model 
with $\CN=4$ supersymmetry. This implies that the resulting manifold is 
Ricci-flat and being four dimensional, it is necessarily hyper-K\"ahler 
\cite{HullAT,9803230}.

\setcounter{equation}{0}
\section{Duality between 6D, $\CN=(1,0)$ CC and CNM sigma-models}

One of the very interesting properties of the nonminimal superfield in four, 
and lower dimensions, is that it is dual to the chiral
multiplet \cite{SUPERSPACE}. In \cite{0508187} we proved that, in flat target
spaces, an analogous duality exists between 6D, $\CN=(1,0)$ CC and CNM 
hypermultiplets. The same happens for 5D sigma--models \cite{projective5d} .
In this section we address the issue of duality for 6D, $\CN=(1,0)$ 
nonlinear sigma--models.

We start by considering the most general CNM sigma--model (\ref{SCNMg}). To 
build its dual we implement the CNM constraint (\ref{constrCNMg2}) using a 
lagrangian multiplier. We then consider the action
\bea
S &=&
\int d^6x d^4\te\,
G\Big(\F^a,\Fib^{\Bar{a}},\Sigma^k,\Bar{\Sigma}^{\Bar{k}}\Big)+
\int d^6x d^2\te\, P_a\Big(\F^b\Big)\pa\,\F^a+
\int d^6x d^2\teb\,\Bar{P}_{\bar{a}}
\Big(\Fib^{\Bar{b}}\Big)\pab\,\Fib^{\Bar{a}}
\non\\
&&~~-\int d^6x d^4\te\Bigg[\,Y_k\Big(\Db^2\S^k-S^k_{a}(\F)\,\pa\,\F^{a}\Big)+
\Bar{Y}_{\Bar{k}}\Big(D^2\Sb^{\Bar{k}}-\Bar{S}^{\Bar{k}}_{\Bar{a}}(\Fib)
\,\pab\,\Fib^{\Bar{a}}\Big)\Bigg]
~~~, 
\label{SCNM2}
\eea
where $Y_k$, $\Bar{Y}_{\Bar{k}}$,  $\S^k$ and  $\Sb^{\Bar{k}}$  
are unconstrained complex superfields. 
Integrating out $Y_k$ and $\Bar{Y}_{\Bar{k}}$ we are back to the original CNM 
model (\ref{SCNMg}, \ref{constrCNMg2}). On the other hand, 
varying with respect to $\S^k$ and $\Sb^{\Bar{k}}$ we obtain the equations of
motion ($G_k \equiv \frac{\pa G}{\pa \S^k}$)
\bea
G_k\,=\,\Db^2Y_k&~~~,~~~&G_{\Bar{k}}\,=\,D^2\Bar{Y}_{\Bar{k}}~~~.
\label{eqDaY}
\eea
We can integrate out $\S^k$ and $\Sb^{\Bar{k}}$ defining new (anti)chiral 
superfields $\chi_k\equiv\Db^2Y_k$, 
$\Bar{\chi}_{\Bar{k}}\equiv D^2\Bar{Y}_{\Bar{k}}$ and inverting the equations
(\ref{eqDaY})
\bea
G_k\Big(\F^a,\Fib^{\Bar{a}},\S^k,\Sb^{\Bar{k}}\Big)\,=\,\chi_k
~&\Longrightarrow&~ \S^k=\S^k
\Big(\F^a,\Fib^{\Bar{a}},\chi_k,\Bar{\chi}_{\Bar{k}}\Big)~~~,
\label{transfCNM1}\\
G_{\Bar{k}}\Big(\F^a,\Fib^{\Bar{a}},\S^k,\Sb^{\Bar{k}}\Big)\,=\,
\Bar{\chi}_{\Bar{k}}
~&\Longrightarrow&~ \Sb^{\Bar{k}}=\Sb^{\Bar{k}}
\Big(\F^a,\Fib^{\Bar{a}},\chi_k,\Bar{\chi}_{\Bar{k}}\Big)~~~.
\label{transfCNM2}
\eea
Substituting back into (\ref{SCNM2}) we find the action for the dual CC model
\bea
\int d^6x\Bigg{\{}
\int d^4\te\Bigg[\,
G\Big(\F^a,\Fib^{\Bar{a}},\S^k,\Sb^{\Bar{k}}\Big)-\S^k\chi_k-
\Sb^{\Bar{k}}\Bar{\chi}_{\Bar{k}}
\,\Bigg]\Bigg|_{\S^k=\S^k(\F^a,\Fib^{\Bar{a}},\chi_k,\Bar{\chi}_{\Bar{k}})}
~~~~~~~
\non\\
+\int d^2\te\Bigg[\Big(\chi_kS^k_{a}+P_a\Big)\pa\,\F^a\,\Bigg]
+\int d^2\teb\Bigg[\Big(\Bar{\chi}_{\Bar{k}}\Bar{S}^{\Bar{k}}_{\Bar{a}}+
\Bar{P}_{\Bar{a}}\Big)\pab\,\Fib^{\Bar{a}}\Bigg]
\Bigg{\}}
~~~~~~~
\non\\
\equiv \int d^6x\Bigg{\{}
\int d^4\te\,\widetilde{G}\Big(\Psi^I,\Bar{\Psi}^{\Bar{I}}\Big)+
\int d^2\te\,Q_I\Big(\Psi^J\Big)\pa\,\Psi^I+
\int d^2\teb\,\Bar{Q}_{\Bar{I}}\Big(\Bar{\Psi}^{\Bar{J}}\Big)
\pab\,\Bar{\Psi}^{\Bar{I}}
\,\Bigg{\}}~~~.~
\label{dual1}
\eea
where we have defined the (anti)chiral superfields 
($\Bar{\Psi}^{\Bar{I}}$) $\Psi^I$ 
\beq
\Psi^I=\(\begin{array}{c} \F^{a}\\
\chi_{k}\end{array}\)~~~,~~~
\Bar{\Psi}^{\Bar{I}}=
\(\begin{array}{c} \Bar{\F}^{\Bar{a}}\\
\Bar{\chi}_{\Bar{k}}\end{array}\)~~~,
\label{defC}
\eeq
being the coordinates of the dual target space. 
The K\"ahler potential $\widetilde{G}$ is
the Legendre transform of $G$ in 
(\ref{SCNMg}) and the (anti)holomorphic pieces are expressed in terms of
\beq
Q_I \equiv \(\begin{array}{c} \chi_kS^k_{a}+P_a\vspace{1ex}\\
0\end{array}\)~~~,~~~
\Bar{Q}_{\Bar{I}} \equiv
\(\begin{array}{c} \Bar{\chi}_{\Bar{k}}\Bar{S}^{\Bar{k}}_{\Bar{a}}
+\Bar{P}_{\Bar{a}}\vspace{1ex}\\
0\end{array}\)~~~.
\label{defQcnmdual}
\eeq
This procedure is very general and allows to map any CNM sigma--model 
(\ref{SCNMg}, \ref{constrCNMg2}) to a CC sigma--model 
(\ref{dual1}--\ref{defQcnmdual}).

A very interesting subclass of dual CC--CNM pairs are those coming from 
projective superspace. Using the prescription just described,
given the CNM sigma--model (\ref{CNMsm}) we can find
the corresponding on--shell 6D, $\CN=(1,0)$ CC sigma--model.
Since in the projective case of section 3 the
CNM multiplet is naturally interpreted as parametrizing the tangent bundle 
$T{\cal M}$ of the K\"ahler manifold ${\cal M}$,
once we perform a duality transformation, the resulting CC coordinates 
$(\F^a,\Fib^{\Bar{a}},\chi_a,\Bar{\chi}_{\Bar{a}})$ describe
the cotangent bundle $T^*{\cal M}$ of ${\cal M}$.
These manifolds must be hyper-K\"ahler as requested from 
the general analysis of the CC case (see section 2). 

It is important to note that in the projective case as well in the free case,
the holomorphic term has the particular form ($P=0$, $S=1$)
\bea
\int d^2\te\,\chi_a\pa\, \F^a\,=\, 
{1\over 2}\int d^2\te \(\Psi^J\O^0_{JI}\)\pa\Psi^I~~~,
\label{sup0}
\eea
where
\bea
\O^0_{IJ}=\(\begin{array}{cc}0&-\d^r_a\\\d^b_k&0\end{array}\)~~~,
\label{symplectic0}
\eea
is the constant symplectic matrix.

The holomorphic term appearing in (\ref{dual1})  
from the CNM$\to$CC dualization is at most linear
in the dualized (anti)chiral superfield ($\Bar{\chi}_{\Bar{k}}$) $\chi_k$. 
At a first sight this term seems
to describe only a subclass of models and one may wonder whether the duality
map does indeed generate the entire class of CC sigma--models.
To answer this question we now prove that performing a suitable change of
coordinates, {\em any} holomorphic term in (\ref{kahler6DCC}) can be always 
reduced locally to 
the canonical form (\ref{sup0}). Therefore, we can state that the duality map
described above is the most general one and relates the whole class of CNM 
models (\ref{SCNMg}, \ref{constrCNMg2}) to the whole class of CC models 
(\ref{kahler6DCC}).

To this end we consider the most general CC sigma--model (\ref{kahler6DCC}) and
search for a holomorphic change of coordinates
\bea
\Psi'^a(\Psi)=f^a(\Psi)&,&\Bar{\Psi}'^{\Bar{a}}(\Bar{\Psi})=
\Bar{f}^{\Bar{a}}(\Bar{\Psi})~~~,\\
\Psi^a(\Psi')=(f^{-1})^a(\Psi')&,&
\Bar{\Psi}^{\Bar{a}}(\Bar{\Psi}')=(\Bar{f}^{-1})^{\Bar{a}}(\Bar{\Psi}')~~~,
\eea
such that
\bea
Q_a(\Psi(\Psi'))\,\pa\,\Psi^a(\Psi')& \equiv &
{1\over 2}\Psi'^b\,\O^0_{ba}\,\pa\,\Psi'^a\,+\,
{\pa g(\Psi')\over \pa \Psi'^a}\,\pa\,\Psi'^a~~~,\\
\Bar{Q}_{\Bar{a}}(\Bar{\Psi}(\Bar{\Psi}'))\,\pab\,
\Bar{\Psi}^{\Bar{a}}(\Bar{\Psi}')& \equiv&
{1\over 2}\Bar{\Psi}'^{\Bar{b}}\,\Ob^0_{\Bar{b}\Bar{a}}\,\pab\,
\Bar{\Psi}'^{\Bar{a}}\,+\,
{\pa \Bar{g}(\Bar{\Psi}')\over \pa\Bar{\Psi}'^{\Bar{a}}}\,\pab\,
\Bar{\Psi}'^{\Bar{a}}~~~.
\eea
The terms ${\pa g(\Psi')\over \pa \Psi'^a}\,\pa\,\Psi'^a=
\pa\,g(\Psi')$ and ${\pa \Bar{g}(\Bar{\Psi}')\over\pa\Bar{\Psi}'^{\Bar{a}}}
\,\pab\,\Bar{\Psi}'^{\Bar{a}}=\pab\,\Bar{g}(\Bar{\Psi}')$, being 
total derivatives, do not affect the holomorphic term and can be always 
admitted in a change of coordinates.

The previous equations are equivalent to the following
differential equations for the functions $f^a$ and $\Bar{f}^{\Bar{a}}$ 
\bea
{1\over 2}\,f^c\,\O^0_{cb}\,{\pa f^b\over\pa\Psi^a}\,+\,
{\pa g\over \pa \Psi^a}&=&Q_a~~~,\\
{1\over 2}\,\Bar{f}^{\Bar{c}}\,\Ob^0_{\Bar{c}\Bar{b}}\,
{\pa \Bar{f}^{\Bar{b}}\over \pa\Bar{\Psi}^{\Bar{a}}}\,+\,
{\pa \Bar{g}\over \pa\Bar{\Psi}^{\Bar{a}}}&=&\Bar{Q}_{\Bar{a}}~~~,
\eea
and also
\bea
\Big(Q_{d(c)}-Q_{c(d)}\Big){\pa \Psi^c\over \pa \Psi'^a}
{\pa \Psi^d\over \pa \Psi'^b}\,=\,
\O_{cd}\,{\pa \Psi^c\over \pa \Psi'^a}{\pa \Psi^d\over \pa \Psi'^b}
&=&\O^0_{ab}~~~,\\
\Big(\Bar{Q}_{\Bar{d}(\Bar{c})}-\Bar{Q}_{\Bar{c}(\Bar{d})}\Big)
{\pa\Bar{\Psi}^{\Bar{c}}\over \pa\Bar{\Psi}'^{\Bar{a}}}
{\pa \Bar{\Psi}^{\Bar{d}}\over \pa\Bar{\Psi}'^{\Bar{b}}}\,= \,
\Ob_{\Bar{c}\Bar{d}}\,{\pa\Bar{\Psi}^{\Bar{c}}\over \pa\Bar{\Psi}'^{\Bar{a}}}
{\pa \Bar{\Psi}^{\Bar{d}}\over \pa\Bar{\Psi}'^{\Bar{b}}}
&=&
\Ob^0_{\Bar{a}\Bar{b}}~~~,
\eea
where the components of the holomorphic two--form $\O$ and $\Ob$
(\ref{OObHyper}) of the hyper-K\"ahler manifold appear.
The functions $Q_a$ transform as the components of a holomorphic one--form 
$Q=Q_a\,d\Psi^a$ and the local change of coordinates described by the previous 
equations is such that the closed, nondegenerate, covariantly constant 
two--form $\O = -\pa\,Q$ is mapped to 
the canonical constant symplectic two--form $\O^0$.
The hyper-K\"ahler manifold is a complex symplectic 
manifold with respect to the holomorphic two--form $\O$. According to 
Darboux theorem we can always choose a particular system of 
coordinates\footnote{See section five of \cite{0512164} for an interesting 
discussion on Darboux coordinates in the case of \\ $~~~~~$ 
generalized K\"ahler geometry. In their language, our (anti)holomorphic 
two-forms $\Ob$ and $\O$ \\ $~~~~~$ 
are those which define the inverse of a Poisson structure on the manifold.}
\cite{Arnold,HuKaLiRo,0512164} for which $\O=\O^0$ and $\Ob=\Ob^0$. This 
insures that locally our previous equations are always soluble.

The discussion above means that locally the 6D CC sigma--models can be always 
described, after the appropriate change of coordinates,
by a holomorphic term having the canonical form (\ref{sup0}).
In the symplectic coordinates it is natural to divide the coordinates as
in (\ref{defC}) and all the CC hyper-K\"ahler sigma--models in the symplectic 
basis reduce to
\beq
S = \int d^6x d^4\te\, K'(\F,\Fib,\chi,\Bar{\chi})+
\int d^6x d^2\te\,\chi_I\,\pa\,\F^I\,+\,
\int d^6x d^2\teb\,\Bar{\chi}_{\Bar{I}}\,\pab\,\Fib^{\Bar{I}}~~~,
\label{CCsym}
\eeq
with $K'$ the K\"ahler potential in the symplectic basis. 

So far we have described how to dualize CNM models obtaining CC 
sigma--models.
Now, we want to proceed in the other way around and construct the CNM dual 
of a general 6D, $\CN=(1,0)$ hyper-K\"ahler CC sigma--model. 
Once we have written it in Darboux coordinates as in (\ref{CCsym}), 
the CC$\to$CNM dualization goes straightforwardly.

We solve the kinematical (anti)chirality constraints of 
($\Bar{\chi}_{\Bar{I}}$) $\chi_I$ in terms of an unconstrained complex 
superfield ($\Bar{Y}_{\Bar{I}}$) $Y_I$ which plays a role similar to the 
Lagrange multiplier of (\ref{dual1})
\bea
\chi_I=-\Db^2 Y_I~~&,&~~\Bar{\chi}_{\Bar{I}}=-D^2\Bar{Y}_{\Bar{I}}~~~.
\eea
The action (\ref{CCsym}) takes the form
\beq
\int d^4\te\Bigg[\,K'(\F,\Fib,\chi,\Bar{\chi})\,-\,
Y_I\,\pa\,\F^I\,-\,\Bar{Y}_{\Bar{I}}\,\pab\,\Fib^{\Bar{I}}\,\Bigg]~~~.
\label{CCsymd}
\eeq
Varying with respect to ($\Bar{Y}_{\Bar{I}}$) $Y_I$ we obtain
\bea
\Db^2\,{\pa K'\over\pa\chi_I}\,=\,\pa\,\F^I~~&,&~~
D^2\,{\pa K'\over\pa\Bar{\chi}_{\Bar{I}}}\,=\,\pab\,\Fib^{\Bar{I}}~~~.
\label{eqchi}
\eea
Therefore, the superfields
\bea
\S^I\,\equiv\,{\pa K'\over\pa\chi_I}~~&,&~~
\Sb^I\,\equiv\,{\pa K'\over\pa\Bar{\chi}_{\Bar{I}}}~~~.
\label{defSSb}
\eea
satisfy the linear constraints in (\ref{constrCNMg2}) with $S=1$. We invert 
these relations to determine  $\chi_I$ and $\Bar{\chi}_{\Bar{I}}$ as functions 
of $(\F, \Fib, \S, \Sb)$. Substituting back into the action (\ref{CCsymd})
we find that the CC sigma--model (\ref{CCsym}) is dual to the CNM sigma--model 
defined by 
\bea
S & = &\int d^6x d^4\te\Bigg[\,
K'\(\F^I,\Fib^{\Bar{I}},\chi_I,\Bar{\chi}_{\Bar{I}}\)\,-\,
\S^I\chi_I\,-\,\Sb^{\Bar{I}}\Bar{\chi}_{\Bar{I}}\,\Bigg]
\Bigg|_{\chi_I=\chi_I(\F^I,\Fib^{\Bar{I}},\S^I,\Sb^{\Bar{I}})}  \non\\
&=&
\int d^6x d^4\te\, \widetilde{K}'\(\F^I,\Fib^{\Bar{I}},\S^I,\Sb^{\Bar{I}}\)~~~,
\eea
where $\widetilde{K}'$ is the Legendre transform with respect to $\Bar{\chi}$ 
and $\chi$ of the K\"ahler potential $K'$.
We have then found that all the CC sigma--models of section 2 written in a 
canonical symplectic system of coordinates are dual to CNM models. 

So far we have considered maximal duality maps, i.e. trasformations
where {\em all} the nonminimal multiplets are dualized to chirals and
viceversa. However, one can consider more general situations where 
the duality map involves only a subset of superfields. These partial
dualizations can be used to map a CNM model with $n_c \neq n_{nm}$ to a 
model with $n_c = n_{nm}$. This is possible every time $n_c - n_{nm} =
2 n$.

On--shell pairs of dual CC--CNM sigma--models have the 
same dynamics. This means that the target space described by the two 
sigma--models is the same. Therefore, on--shell the CNM model describes 
a hyper-K\"ahler manifold as well. 
In particular, as also noted in \cite{9803230}, the duality 
Legendre transform acts on the manifold as a change of 
coordinates which is in general non--holomorphic (not preserving the complex 
structures).

\section{6D, $\CN=(1,0)$ CNM sigma-models (II): an indirect approach from its
dual CC model}

In section 4 we have studied the most general CNM sigma--model defined by
our ansatz (\ref{SCNMg}, \ref{constrCNMg2}) 
and worked out the constraints on its defining functions 
as coming from the direct restoration of 6D Lorentz invariance of the on--shell
action. 
Unfortunately, as already noticed, the system of constraints which we obtain 
cannot be solved in general and we are not able to
easily read from them the geometrical properties of the target space. 

In the previous section we have discussed the duality properties between
CC and CNM sigma models. This opens the possibility to find the set of 
constraints
satisfied by the CNM sigma--model (\ref{SCNMg}, \ref{constrCNMg2}) 
by following
an alternative, indirect approach: Since we know the precise relation between 
the geometric
tensors of the CNM model and of its dual we can infer the constraints of 
the CNM case from the hyper-K\"ahler condition (\ref{constrCC}) for the dual CC
model.

Given the general CNM (\ref{SCNMg}, \ref{constrCNMg2}) we can find
the components of the two--forms $\O$ and $\Ob$ for the CC dual (\ref{dual1})  
\bea
\O_{IJ}&=&Q_{J(I)}-Q_{I(J)}=
\begin{small}
\(\begin{array}{cc}\Big(P_{b(a)}-P_{a(b)}\Big)+
\chi_s\Big(S^s_{b(a)}-S^s_{a(b)}\Big)
\vspace{1ex}
&~~-S^r_{a}\\
S^k_{b}\vspace{1ex}
&~~0\end{array}\)
\end{small},~~~~~~~
\label{twoFormCNM1}\\\non\\
\Ob_{\Bar{I}\Bar{J}}&=&\Bar{Q}_{\Bar{J}(\Bar{I})}-\Bar{Q}_{\Bar{I}(\Bar{J})}=
\begin{small}
\(\begin{array}{cc}
\Big(\Bar{P}_{\Bar{b}(\Bar{a})}-\Bar{P}_{\Bar{a}(\Bar{b})}\Big)+
\Bar{\chi}_{\Bar{s}}\Big(\Bar{S}^{\Bar{s}}_{\Bar{b}(\Bar{a})}-
\Bar{S}^{\Bar{s}}_{\Bar{a}(\Bar{b})}\Big)
\vspace{1ex}
&~-\Bar{S}^{\Bar{r}}_{\Bar{a}}\\
\Bar{S}^{\Bar{k}}_{\Bar{b}}
\vspace{1ex}
&~0\end{array}\)
\end{small}
.~~~~~~~
\label{twoFormCNM2}
\eea
To write the hyper-K\"ahler conditions (\ref{constrCC}) we need find the 
expression of the K\"ahler metric 
$\widetilde{G}_{I\Bar{I}}=\pa_I\pa_{\Bar{I}}\,\widetilde{G}$ of the dual CC 
geometry in terms of the tensors in the CNM basis. Exploiting the fact that
the K\"ahler potential is the Legendre transform of $G$, we find
\bea
\widetilde{G}_{a\Bar{a}}&=&\pa_{\Bar{a}}\Bigg[G_{a}+
G_{k}{\pa \S^k\over\pa \F^a}+
G_{\Bar{k}}{\pa \Sb^{\Bar{k}}\over\pa \F^a}-
\chi_k{\pa \S^k\over\pa \F^a}-
\Bar{\chi}_{\Bar{k}}{\pa \Sb^{\Bar{k}}\over\pa \F^a}
\Bigg] \non\\
&=&G_{a\Bar{a}}+G_{ak}{\pa \S^k\over\pa \Fib^{\Bar{a}}}
+G_{a\Bar{k}}{\pa \Sb^{\Bar{k}}\over\pa \Fib^{\Bar{a}}}~=~
G_{a\Bar{a}}+G_{\Bar{a}k}{\pa \S^k\over\pa \F^a}
+G_{\Bar{a}\Bar{k}}{\pa \Sb^{\Bar{k}}\over\pa \F^a}~~~,\\
\widetilde{G}^k_{~\Bar{a}}&=&\pa^kG_{\Bar{a}}~=~
G_{r\Bar{a}}{\pa \S^r\over\pa \chi_k}+
G_{\Bar{a}\Bar{r}}{\pa \Sb^{\Bar{r}}\over\pa \chi_k} \non\\
&=&\pa_{\Bar{a}}\Bigg[G_{r}{\pa \S^r\over\pa \chi_k}
+G_{\Bar{r}}{\pa \Sb^r\over\pa \chi_k}-
{\pa \S^r\over\pa \chi_k}\chi_r-\S^k-
{\pa \Sb^r\over\pa \chi_k}\Bar{\chi}_{\Bar{r}}\Bigg]~=~
-{\pa \S^k\over\pa \Fib^{\Bar{a}}}~~~,\\
\widetilde{G}_a^{~\Bar{k}}&=&
G_{ar}{\pa \S^r\over\pa \Bar{\chi}_{\Bar{k}}}+
G_{a\Bar{r}}{\pa \Sb^{\Bar{r}}\over\pa \Bar{\chi}_{\Bar{k}}}
~=~-{\pa \Sb^{\Bar{k}}\over\pa \F^{a}}~~~,\\
\widetilde{G}^{k\Bar{k}}&=&-{\pa \S^k\over\pa \Bar{\chi}_{\Bar{k}}}~=~
-{\pa \Sb^{\Bar{k}}\over\pa \chi_k}~~~.
\eea
Using the defining equations (\ref{transfCNM1}, \ref{transfCNM2}) of the CNM
Legendre transform we have 
\bea
H\,=\,
\(\begin{array}{cc}G_{kr}\vspace{1ex}&G_{k\Bar{r}}\\
G_{\Bar{k}r}\vspace{1ex}&G_{\Bar{k}\Bar{r}}\end{array}\)\,=\,
\(\begin{array}{cc}{\pa \chi_k\over\pa\S^r}\vspace{1ex}&
{\pa \chi_k\over\pa \Sb^{\Bar{r}}}\vspace{1ex}\\
{\pa\Bar{\chi}_{\Bar{k}}\over\pa\S^r}&
{\pa \Bar{\chi}_{\Bar{k}}\over\pa \Sb^{\Bar{r}}}\end{array}\)~~~,
\label{2H}\\
H^{-1}\,\equiv\,
\(\begin{array}{cc}H^{kr}\vspace{1ex}&H^{k\Bar{r}}\\
H^{\Bar{k}r}\vspace{1ex}&H^{\Bar{k}\Bar{r}}\end{array}\)\,=\,
\(\begin{array}{cc}{\pa\S^k\over\pa \chi_r}\vspace{1ex}&
{\pa\S^k\over\pa \Bar{\chi}_{\Bar{r}}}\\
{\pa \Sb^{\Bar{k}}\over\pa \chi_r}\vspace{1ex}&
{\pa \Sb^{\Bar{k}}\over\pa \Bar{\chi}_{\Bar{k}}}\end{array}\)~~~.
\label{H-1}
\eea
We are then able to write the metric of the CC K\"ahler target space in terms 
of tensors of the original CNM model
\bea
\widetilde{G}_{I\Bar{J}}&=&
\(\begin{array}{cc}\widetilde{G}_{a\Bar{b}}\vspace{1ex}&
\widetilde{G}_a^{~\Bar{r}}\\
\widetilde{G}^k_{~\Bar{b}}\vspace{1ex}&\widetilde{G}^{k\Bar{r}}\end{array}\)
~~~,
\label{KdualCNM}
\eea
with
\bea
\widetilde{G}_{a\Bar{a}}&=&
G_{a\Bar{a}}-G_{ak}(H^{kr}G_{r\Bar{a}}+
H^{k\Bar{r}}G_{\Bar{r}\Bar{a}})-
G_{a\Bar{k}}(H^{\Bar{k}r}G_{r\Bar{a}}+H^{\Bar{k}\Bar{r}}G_{\Bar{r}\Bar{a}})~~~,
\label{KdualCNM1}\\
\widetilde{G}_a^{~\Bar{k}}&=&
G_{ar}H^{r\Bar{k}}+G_{a\Bar{r}}H^{\Bar{r}\Bar{k}}~~~,\label{KdualCNM2}\\
\widetilde{G}^k_{~\Bar{a}}&=&
G_{r\Bar{a}}H^{rk}+G_{\Bar{a}\Bar{r}}H^{\Bar{r}k}~~~,\label{KdualCNM3}\\
\widetilde{G}^{k\Bar{k}}&=&-H^{k\Bar{k}}~~~.
\eea
In a matricial form as (\ref{defMatr2}, \ref{defMatr3}) and 
(\ref{2H}, \ref{H-1}), the dual CC K\"ahler metric (\ref{KdualCNM}) can be 
written as
\bea
\widetilde{G}\,=\,P_+\(M\,-\,NH^{-1}N^TP_-\)P_2\,-\,
{1\over 4}\,P_2[P,[P,H^{-1}]]P_-\,+\,
{1\over 4}\,[P,[P,NH^{-1}]]~~~,~~~~
\label{KdualCNM4}
\eea
where
\bea
P_2&\equiv&
\(\begin{array}{cc}0&1\\1&0\end{array}\)~~~.
\eea

At this point we impose the hyper-K\"ahler condition (\ref{constrCC})
\bea
\widetilde{G}_{I\Bar{I}}\,=\,-\,\O_{IJ}\,
\widetilde{G}^{J\Bar{J}}\,\Ob_{\Bar{J}\Bar{I}}~~~\Longleftrightarrow~~~
\widetilde{G}_{I\Bar{I}}\,\O^{IJ}\,\widetilde{G}_{J\Bar{J}}=\,-\,
\Ob_{\Bar{I}\Bar{J}}~~~.
\label{HyperCNMCC}
\eea
These equations can be re-interpreted as conditions on the CNM
tensors, remembering that $\chi_k=G_k(\F,\Fib,\S,\Sb)$.
Therefore, inverting the metric $\widetilde{G}_{I\Bar{J}}$ or the
form $\O_{IJ}$ one can write explicitly the conditions on the tensors of the
original CNM sigma--model which insures on--shell 6D 
Lorentz and $\CN=(1,0)$ SUSY on both sides of the duality map.

We have not investigated extensively the constraints (\ref{HyperCNMCC}) 
for the generic sigma--model (\ref{SCNMg}, \ref{constrCNMg2}) yet. 
However, we can make few preliminary observations. 

So far, we have not specified the number
of coordinates of the target space described respectively by chiral $\F^a$ and
nonminimal $\S^k$ (or dual chiral $\chi_k$) superfields. In order to understand
if there are restrictions on the number of coordinates we analyse
the tensors $\O_{IJ}$, $\Ob_{\Bar{I}\Bar{J}}$ 
(\ref{twoFormCNM1}, \ref{twoFormCNM2}). To have a well--defined dual CC model 
with on--shell 6D, $\CN=(1,0)$ SUSY we know that $\O_{IJ}$
has to provide a local parametrization of the components of the 
nondegenerate closed holomorphic two--form of a hyper-K\"ahler manifold. 
In particular, the matrix $\O_{IJ}$ has to be invertible, i.e. its kernel has 
to be trivial. 
Observing the explicit expression of $\O_{IJ}$ (\ref{twoFormCNM1}) in the 
case under consideration it is clear that the number of coordinates described
by nonminimal superfields $n_{nm}$ has to be equal or less as the number of
chirals $n_c$ ($n_{nm}\le n_c$). In fact, if $n_{nm}>n_c$ then $S^k_a$ would
certainly have a non--trivial kernel and so would $\O_{IJ}$ 
(\ref{twoFormCNM1}).

We first consider the case $n_{nm}= n_c \equiv n$.
Since $\O_{IJ}$ has to be invertible, we require
$S^k_{a}(\F)$ to have a trivial kernel and an inverse
$S^a_{k}(\F)$ exists such that $S^k_{b}S^b_{r}=\d^k_r$, 
$S^a_{r}S^r_{b}=\d^a_b$.
If $S^k_{a}$ is invertible, it is possible to simplify the CC dual 
(\ref{dual1}) by doing the  holomorphic affine--like $\chi_k$, 
$\Bar{\chi}_{\Bar{k}}$ superfield redefinition
\bea
\tilde{\chi}_k\,\equiv\,\chi_r\,S^r_a(\F)\,\d^a_k\,+\,P_{a}(\F)\,\d^a_k&,&
\Bar{\tilde{\chi}}_{\Bar{k}}\,\equiv\,
\Bar{\chi}_{\Bar{r}}\,\Bar{S}^{\Bar{r}}_{\Bar{a}}(\Fib)\,\d^{\Bar{a}}_{\Bar{k}}
\,+\,\Bar{P}_{\Bar{a}}(\Fib)\,\d^{\Bar{a}}_{\Bar{k}}~~~,\\
\chi_k\,=\,\tilde{\chi}_r\,\d^r_b\,S^b_k(\F)\,-\,P_{a}(\F)\,\d^a_k&,&
\Bar{\chi}_{\Bar{k}}\,=\,\Bar{\tilde{\chi}}_{\Bar{r}}\,\d^{\Bar{r}}_{\Bar{b}}\,
\Bar{S}^{\Bar{b}}_{\Bar{r}}(\Fib)\,-\,
\Bar{P}_{\Bar{a}}(\Fib)\,\d^{\Bar{a}}_{\Bar{k}}~~~,
\eea
keeping the $\F$, $\Fib$ coordinates fixed. In the 
$(\F,\Fib,\tilde{\chi}_k, \Bar{\tilde{\chi}}_{\Bar{k}})$
target space coordinates the CC sigma--model which we find is in a symplectic 
basis where the holomorphic term is (\ref{sup0}). If we now 
dualize the resulting CC sigma--model with respect to the new tilde 
coordinates, we find a CNM sigma--model where $P_a\equiv 0$ and 
$S^k_a(\F)\equiv\d^k_a$. This means that, with $n_{nm}=n_c$ all the consistent 
6D, $\CN=(1,0)$ CNM sigma--models can be described by $P_a=0$ and 
$S^k_a(\F)=\d^k_a$. 
This is clearly what we expect from the discussion of section 5. We then 
focus on this particular case.

With $P_a=0$ and $S^k_a=\d^k_a$, $\O_{IJ}$ and $\O^{IJ}$ 
become the constant symplectic matrix and its inverse, respectively 
(also $\O_{IJ}=P_2P$ and $\O^{IJ}=-P_2P=PP_2$).  
The hyper-K\"ahler condition on the CC dual sigma--model then reduces to
\bea
0&=&H^{k\Bar{k}}\d^a_{k}(G_{ap}H^{p\Bar{r}}\,+\,
G_{a\Bar{p}}H^{\Bar{p}\Bar{r}})\,-\,
H^{k\Bar{r}}\d^a_{k}(G_{ap}H^{p\Bar{k}}\,+\,
G_{a\Bar{p}}H^{\Bar{p}\Bar{k}})~~~,\\
-\,\d^{\Bar{k}}_{\Bar{b}}&=&
H^{k\Bar{k}}\d^b_{k}\Big[G_{b\Bar{b}}\,
-G_{bs}(H^{sr}G_{r\Bar{b}}+H^{s\Bar{r}}G_{\Bar{r}\Bar{b}})-
G_{b\Bar{s}}(H^{\Bar{s}r}G_{r\Bar{b}}+H^{\Bar{s}\Bar{r}}G_{\Bar{r}\Bar{b}})
\Big] \non\\
&&+\,(G_{as}H^{s\Bar{k}}\,+\,G_{a\Bar{s}}H^{\Bar{s}\Bar{k}})\d^a_{k}
(G_{\Bar{b}p}H^{pk}\,+\,G_{\Bar{b}\Bar{p}}H^{\Bar{p}k})~~~,\\
0&=&\Bigg[\,G_{a\Bar{a}}-G_{as}(H^{sr}G_{r\Bar{a}}+
H^{s\Bar{r}}G_{\Bar{r}\Bar{a}}) \non\\
&&~~~~~~~~~
-G_{a\Bar{s}}(H^{\Bar{s}r}G_{r\Bar{a}}+H^{\Bar{s}\Bar{r}}G_{\Bar{r}\Bar{a}})
\,\Bigg]
\d^a_{k}(G_{\Bar{b}p}H^{pk}\,+\,G_{\Bar{b}\Bar{p}}H^{\Bar{p}k}) \non\\
&&-\,\Bigg[\,G_{a\Bar{b}}-G_{as}(H^{sr}G_{r\Bar{b}}+
H^{s\Bar{r}}G_{\Bar{r}\Bar{b}}) \non\\
&&~~~~~~~~~~~\,
-\,G_{a\Bar{s}}(H^{\Bar{s}r}G_{r\Bar{b}}+H^{\Bar{s}\Bar{r}}G_{\Bar{r}\Bar{b}})
\,\Bigg]
\d^a_{k}(G_{\Bar{a}p}H^{pk}\,+\,G_{\Bar{a}\Bar{p}}H^{\Bar{p}k})~~~,~~~
\eea
or in the matricial form 
\bea
P_2P&=&\widetilde{G}^T\,P_2P\,\widetilde{G}~~~.
\eea
where $\widetilde{G}$ is given by (\ref{KdualCNM4}).

If $n_{nm}<n_c$ and $n_{nm}+n_c=2n$, we expect that under a set of partial
dualities the CNM model can be 
mapped to a CNM model with $n_{nm}=n_c$, as discussed in section 5.
Therefore, the previous analysis still works.
On the other hand, if $n_{nm}+n_c=2n+1$ the theory is not well-defined since 
an odd number of target space coordinates is incompatible with the
hyper-K\"ahler condition.

\paragraph{Example: 4D target space}

We now analyse the 4D target space example of section 4 
(see the action (\ref{Target4DCNM})) using the 
indirect approach of this section. The dual CC K\"ahler metric 
$\widetilde{G}_{I\Bar{J}}$ (\ref{KdualCNM}) is 
\bea
\widetilde{G}_{\F\Fib}=
G_{\F\Fib}-\frac{1}{\det{H}}
\Big[G_{\F\S}(G_{\Sb\Sb}G_{\S\Fib}-G_{\S\Sb}G_{\Fib\Sb})+
G_{\F\Sb}(G_{\S\S}G_{\Fib\Sb}-G_{\S\Sb}G_{\S\Fib})
\Big]~,~~~\\
\widetilde{G}_\F^{\,~\Sb}=
\frac{1}{\det{H}}\Big[G_{\F\Sb}G_{\S\S}-G_{\F\S}G_{\S\Sb}\Big]~,
~~~~~~~~~~~~~~~~~~~~~~~~~~~~~~~~~~~~~~~~~~~~~~~~~~~~~~~~~\,\,\\
\widetilde{G}^\S_{~\,\Fib}=
\frac{1}{\det{H}}\Big[G_{\S\Fib}G_{\Sb\Sb}-G_{\Fib\Sb}G_{\S\Sb}\Big]~,
~~~~~~~~~~~~~~~~~~~~~~~~~~~~~~~~~~~~~~~~~~~~~~~~~~~~~~~~~\,\,\\
\widetilde{G}^{\S\Sb}=
\frac{1}{\det{H}}\,G_{\S\Sb~}~.~~
~~~~~~~~~~~~~~~~~~~~~~~~~~~~~~~~~~~~~~~
~~~~~~~~~~~~~~~~~~~~~~~~~~~~~~~~~~~~~~~\,
\eea

Imposing the hyper-K\"ahler condition 
(\ref{HyperCNMCC}) with 
$\O_{IJ}=\O^0_{IJ}=\begin{footnotesize}
\(\begin{array}{cc}0&-1\\1&0\end{array}\)\end{footnotesize}$, the only 
constraint on $G(\F,\Fib,\S,\Sb)$ which arises is 
(\ref{constrTarget4DCNM}). This proves that, at least in the case of a four 
dimensional target space geometry, our direct and indirect approaches are 
equivalent.
Furthermore, from the present discussion it follows that  
(\ref{constrTarget4DCNM}) is effectively a hyper-K\"ahler condition as we 
claimed at the end of section 4.

\setcounter{equation}{0}
\section{Conclusions}

In this work, we have endeavored to open a discussion of 6D supersymmetric
nonlinear sigma-models.  Our specific techniques involved utilizing 4D
superfields, thus keeping manifest this degree of supersymmetry, that permit
full 6D Lorentz invariance to be realized only on-shell.  

We have demonstrated, as might have been expected, that the use of two
chiral superfields to represent the 6D $\cal N$ $=$ $(1,\,0)$ hypermultiplet
provides the simplest manner in which to describe such actions.  
This formulation has the interesting feature that to write its action requires 
in addition to a K\" ahler
potential, a holomorphic U$(1)$-bundle connection.  The 6D Lorentz invariance 
imposes
a condition that relates the K\" ahler potential to the connection in such a
way that the sigma-model manifold must be hyper-K\" ahler.  The field strength
of the holomorphic U$(1)$-bundle connection has been found to be related to the
well-known triplet of covariantly constant complex structures.   We have also 
given a brief introduction to the use of projective superspace for analysis of 
this class of models.  
As the polar multiplets of projective superspace necessarily lead to 
combinations
of chiral and nonmininal multiplets (CNM's) making their appearance, we finally
have studied this class of models by an analysis based directly on the 
introduction of CNM actions without the use of projective superspace.  In this 
last set of activities, general conditions were derived, but owing to the sheer
algebraic complexity, we have shown that there exist well define special cases 
which demonstrate the equivalence of the CNM description, where possible.

\section*{Acknowledgements}

We would like to thank E. A. Ivanov, S. V. Ketov, S. M. Kuzenko and M. 
Ro\v cek for comments.
G. T.-M. would like to thank the Department of Physics for hospitality during 
the initial stage of this work. As well we acknowledge partial support from the
Center for String and Particle Theory of the University of Maryland.
S. J. G. is supported in part by National Science Foundation Grant PHY-0354401.
S. P. and G. T.-M. are partially supported by INFN, PRIN prot. 
$2005-024045-004$ and the European Commission RTN program 
MRTN--CT--2004--005104.

\appendix

\setcounter{equation}{0}
\section{Some definitions and formulae}

In analogy to the four dimensional case \cite{DeoGates,CNMmassless,0404222} 
we define the component fields of the CNM multiplet 
(\ref{constrCNMg2}) as 
\bea
&A^{a}=\F^{a}|~~~,~~~\psi^{a}_\a=D_\a\F^{a}|~~~,~~~F^{a}=D^2\F^{a}|~~~,\non\\
&\Bar{A}^{\Bar{a}}=\Fib^{\Bar{a}}|~~~,~~~
\Bar{\psi}^{\Bar{a}}_\ad=\Db_\ad\Fib^{\Bar{a}}|~~~,~~~
\Bar{F}^{\Bar{a}}=\Db^2\Fib^{\Bar{a}}|~~~,\non\\
&B^{k}=\S^k|~~~,~~~\Bar{\z}^{k}_\ad=\Db_\ad\S^{k}|~~~,~~~H^{k}=D^2\S^{k}|~~~,
\non\\
&\rho^{k}_\a=D_\a\S^{k}|~~~,~~~p^{k}_{\a\ad}=\Db_\ad D_\a\S^{k}|~~~,~~~
\Bar{\b}^{k}_\ad={1\over 2}D^\a\Db_\ad D_\a\S^{k}|~~~.
\non\\
&\Bar{B}^{\Bar{k}}=\Sb^{\Bar{k}}|~~~,~~~
\z^{\Bar{k}}_\a=D_\a\Sb^{\Bar{k}}|~~~,~~~
\Bar{H}^{\Bar{k}}=\Db^2\Sb^{\Bar{k}}|~~~,\non\\
&\Bar{\rho}^{\Bar{k}}_\ad=\Db_\ad\Sb^{\Bar{k}}|~~~,~~~
\Bar{p}^{\Bar{k}}_{\a\ad}=-D_\a \Db_\ad\Sb^{k}|~~~,~~~
\b^{\Bar{k}}_\a={1\over 2}\Db^\ad D_\a \Db_\ad\Sb^{\Bar{k}}|~~~.
\label{componentsCNM}
\eea
From the bosonic components we define the vectors
\bea
{\cal A}\equiv\(\begin{array}{c} A^a\\
\Bar{A}^{\Bar{a}}\end{array}\)~~~,~~~
{\cal B}=\(\begin{array}{c} B^k\\\Bar{B}^{\Bar{k}}\end{array}\)
~~~,~~~~~~~~~~~~~\\
{\cal F}\equiv\(\begin{array}{c} F^a\\
\Bar{F}^{\Bar{a}}\end{array}\)~~~,~~~
{\cal H}\equiv\(\begin{array}{c} H^k\\
\Bar{H}^{\Bar{k}}\end{array}\)~~~,~~~
{\cal P}_{\a\ad}\equiv\(\begin{array}{c} p^k_{\a\ad}\\
\Bar{p}^{\Bar{k}}_{\a\ad}\end{array}\)~~~.
\label{defVectors}
\eea

\setcounter{equation}{0}
\section{6D CNM sigma--model action in components}

Now we give the expression of the action in components for the general
CNM sigma--model described by (\ref{SCNMg}) with constraints 
(\ref{constrCNMg2}). By non-trivial dimensional reduction we can obtain  
component actions in lower dimensions. In particular, the sigma--model actions
we obtain can contain non--trivial mass and potential terms 
coming from the CNM constraint. These actions generalize sigma--models studied 
in \cite{DeoGates,CNMmassless} where only the standard nonminimal constraint
$\Bar{D}^2 \S = 0$ was considered.
Our more general models are relevant for a CNM description of SUSY theories 
with non-trivial central charges.

With the components defined as (\ref{componentsCNM}), the action of the CNM 
sigma--model (\ref{SCNMg}, \ref{constrCNMg2}) is
\bea
&&\Big(P_{b(a)}-P_{a(b)}\Big)\Big(F^a\pa A^b+
{1\over 2}\psi^{b\a}\pa\,\psi^a_\a\Big)
+{1\over 2}\Big(P_{b(ac)}-P_{a(bc)}\Big)(\pa A^a)\,\psi^{b\a}\psi^c_\a
\non\\
&&+\,\Big(\Bar{P}_{\Bar{b}(\Bar{a})}-\Bar{P}_{\Bar{a}(\Bar{b})}\Big)
\Big(\Bar{F}^{\Bar{a}}\pab\,\Bar{A}^{\Bar{b}}+
{1\over 2}\Bar{\psi}^{\Bar{b}\ad}\pab\,\Bar{\psi}^{\Bar{a}}_\ad\Big)
+{1\over 2}\Big(\Bar{P}_{\Bar{b}\,(\Bar{a}\Bar{c})}-
\Bar{P}_{\Bar{a}\,(\Bar{b}\Bar{c})}\Big)(\pab\,\Bar{A}^{\Bar{a}})\,
\Bar{\psi}^{\Bar{b}\ad}\Bar{\psi}^{\Bar{c}}_\ad
\non\\
&&+\,
G_k S^k_{a(b)}F^b\pa\, A^a
+G_k S^k_a\pa F^a+
G_k S^k_{a(b)}\psi^{b\a}\pa\,\psi^a_\a+
G_kS^k_{a(bc)}(\pa A^a)\psi^{b\a}\psi^c_\a
\non\\
&&+\,\Bar{G}_{\Bar{k}}\Bar{S}^{\Bar{k}}_{\Bar{a}(\Bar{b})}\Bar{F}^{\Bar{b}}
\pab\,\Bar{A}^{\Bar{a}}+
\Bar{G}_{\Bar{k}}\Bar{S}^{\Bar{k}}_{\Bar{a}}\pab\,\Bar{F}^{\Bar{a}}+
\Bar{G}_{\Bar{k}}\Bar{S}^{\Bar{k}}_{\Bar{a}(\Bar{b})}
\Bar{\psi}^{\Bar{b}\ad}\pab\,\Bar{\psi}^{\Bar{a}}_\ad
+\,\Bar{G}_{\Bar{k}}\Bar{S}^{\Bar{k}}_{\Bar{a}(\Bar{b}\Bar{c})}
(\pab\,\Bar{A}^{\Bar{a}})\Bar{\psi}^{\Bar{b}\ad}\Bar{\psi}^{\Bar{c}}_\ad
\non\\
&&+\,G_{ak}\Bigg[\,S^k_{b}F^a\pa\, A^b+S^k_{b}\psi^{a\a}\pa\,\psi^b_\a+
S^k_{b(c)}(\pa\,A^b)\psi^{a\a}\psi^c_\a
\,\Bigg]
\non\\
&&+\,G_{\Bar{a}\Bar{k}}\Bigg[\,\Bar{S}^{\Bar{k}}_{\Bar{b}}\Bar{F}^{\Bar{a}}
\pab\,\Bar{A}^{\Bar{b}}+
\Bar{S}^{\Bar{k}}_{\Bar{b}}\Bar{\psi}^{\Bar{a}\ad}
\pab\,\Bar{\psi}^{\Bar{b}}_\ad+
\Bar{S}^{\Bar{k}}_{\Bar{b}(\Bar{c})}
(\pab\,\Bar{A}^{\Bar{b}})\Bar{\psi}^{\Bar{a}\ad}\Bar{\psi}^{\Bar{c}}_\ad
\,\Bigg]
\non\\
&&+\,G_{kr}\Bigg[\,S^k_aH^r\pa\,A^a+{1\over 2}\pa^{\a\ad}B^{k}\pa_{\a\ad}B^{r}
+p^{k\a\ad}i\pa_{\a\ad}B^{r}-{1\over 2}p^{k\a\ad}p^{r}_{\a\ad}
-\Bar{\z}^{k\ad}\Bar{\b}^r_\ad
\non\\
&&~~~~~~~~
+\rho^{r\a}\Big({1\over 2}i\pa_{\a\ad}\Bar{\z}^{k\ad}+
S^k_a\pa\,\psi^a_\a+S^k_{a(b)}(\pa\,A^a)\psi^b_\a\Big)\Bigg]
\non\\
&&+\,G_{\Bar{k}\Bar{r}}\Bigg[\,\Bar{S}^{\Bar{k}}_{\Bar{a}}\Bar{H}^{\Bar{r}}
\pab\,\Bar{A}^{\Bar{a}}
+{1\over 2}\pa^{\a\ad}\Bar{B}^{\Bar{k}}\pa_{\a\ad}\Bar{B}^{\Bar{r}}
-\Bar{p}^{\Bar{k}\a\ad}i\pa_{\a\ad}\Bar{B}^{\Bar{r}}
-{1\over 2}\Bar{p}^{\Bar{k}\a\ad}\Bar{p}^{\Bar{r}}_{\a\ad}
-{\z}^{\Bar{k}\a}{\b}^{\Bar{r}}_\a
\non\\
&&~~~~~~~~
+\,\Bar{\rho}^{\Bar{r}\ad}\Big({1\over 2}i\pa_{\a\ad}{\z}^{\Bar{k}\a}+
\Bar{S}^{\Bar{k}}_{\Bar{a}}\pab\,\Bar{\psi}^{\Bar{a}}_\ad+
\Bar{S}^{\Bar{k}}_{\Bar{a}(\Bar{b})}(\pab\,\Bar{A}^{\Bar{a}})
\Bar{\psi}^{\Bar{b}}_\ad\Big)\Bigg]
\non\\
&&+\,G_{a\Bar{a}}\Bigg[-{1\over 2}\pa^{\a\ad}\Bar{A}^{\Bar{a}}\pa_{\a\ad}A^a
-{1\over 2}\Bar{\psi}^{\Bar{a}}_\ad i\pa^{\a\ad}\psi^{a}_\a
-{1\over 2}\psi^{a}_\a i\pa^{\a\ad}\Bar{\psi}^{\Bar{a}}_\ad+
\Bar{F}^{\Bar{a}}F^a\,\Bigg]
\non\\
&&+\,G_{k\Bar{k}}\Bigg[\,{1\over 2}\pa^{\a\ad}\Bar{B}^{\Bar{k}}\pa_{\a\ad}B^k
+{1\over 2}\Bar{\z}^{k}_\ad i\pa^{\a\ad}\z^{\Bar{k}}_\a
+{1\over 2}\z^{\Bar{k}}_\a i\pa^{\a\ad}\Bar{\z}^{k}_\ad
-\Bar{p}^{\Bar{k}\a\ad}p^k_{\a\ad}
+S^k_a\Bar{S}^{\Bar{k}}_{\Bar{a}}(\pab\,\Bar{A}^{\Bar{a}})(\pa\,A^a)
\non\\
&&~~~~~~~~
+\,\Bar{H}^{\Bar{k}}H^k+S^k_a\z^{\Bar{k}\a}\pa\,\psi^a_\a+
S^k_{a(b)}(\pa\,A^a)\z^{\Bar{k}\a}\psi^b_\a-\b^{\Bar{k}\a}\rho^k_\a
+\Bar{S}^{\Bar{k}}_{\Bar{a}}\Bar{\z}^{k\ad}\pab\,\Bar{\psi}^{\Bar{a}}_\ad
\non\\
&&~~~~~~~~
+\,\Bar{S}^{\Bar{k}}_{\Bar{a}(\Bar{b})}(\pab\,\Bar{A}^{\Bar{a}})
\Bar{\z}^{k\ad}\Bar{\psi}^{\Bar{b}}_\ad
-\Bar{\b}^{k\ad}\Bar{\rho}^{\Bar{k}}_\ad\,\Bigg]
\non\\
&&+\,G_{a\Bar{k}}\Bigg[\,{1\over 2}\pa^{\a\ad}\Bar{B}^{\Bar{k}}\pa_{\a\ad}A^a
-\Bar{p}^{\Bar{k}\a\ad}i\pa_{\a\ad}A^a+
\Bar{H}^{\Bar{k}}F^a-\psi^{a\a}\b^{\Bar{k}}_\a
-{1\over 2}\Bar{\rho}^{\Bar{k}}_\ad i\pa^{\a\ad}\psi^{a}_\a
\,\Bigg]
\non\\
&&+\,G_{k\Bar{a}}\Bigg[\,{1\over 2}\pa^{\a\ad}\Bar{A}^{\Bar{a}}\pa_{\a\ad}B^k
+p^{k\a\ad}i\pa_{\a\ad}\Bar{A}^{\Bar{a}}+
\Bar{F}^{\Bar{a}}H^k-\Bar{\psi}^{\Bar{a}\ad}\Bar{\b}^{k}_\ad
-{1\over 2}\rho^k_\a i\pa^{\a\ad}\Bar{\psi}^{\Bar{a}}_\ad
\,\Bigg]
\non\\
&&+\,
G_{ak\Bar{k}}\Bigg[\,{i\over 2}(\pa^{\a\ad}A^a)\z^{\Bar{k}}_\a\Bar{\z}^k_\ad
+{i\over 2}(\pa^{\a\ad}A^a)\rho^{k}_\a\Bar{\rho}^{\Bar{k}}_\ad+
F^a\Bar{\z}^{k\ad}\Bar{\rho}^{\Bar{k}}_\ad
+\Bar{H}^{\Bar{k}}\psi^{a\a}\rho^k_\a
+S^k_b(\pa\,A^b)\psi^{a\a}\z^{\Bar{k}}_\a
\non\\
&&~~~~~~~~~
+\,\Bar{p}^{\Bar{k}\a\ad}\psi^{a}_\a\Bar{\z}^k_\ad
-{i\over 2}(\pa^{\a\ad}B^k)\psi^{a}_\a\Bar{\rho}^{\Bar{k}}_\ad
+p^{k\a\ad}\psi^a_\a\Bar{\rho}^{\Bar{k}}_\ad\,\Bigg]
\non\\
&&+\,G_{k\Bar{a}\Bar{k}}
\Bigg[\,{i\over 2}(\pa^{\a\ad}\Bar{A}^{\Bar{a}})\Bar{\z}^{k}_\ad\z^{\Bar{k}}_\a
+{i\over 2}(\pa^{\a\ad}\Bar{A}^{\Bar{a}})\Bar{\rho}^{\Bar{k}}_\ad\rho^k_\a+
\Bar{F}^{\Bar{a}}\z^{\Bar{k}\a}\rho^k_\a
+H^k\Bar{\psi}^{\Bar{a}\ad}\Bar{\rho}^{\Bar{k}}_\ad+
\Bar{S}^{\Bar{k}}_{\Bar{b}}(\pab\,\Bar{A}^{\Bar{b}})\Bar{\psi}^{\Bar{a}\ad}
\Bar{\z}^k_\ad
\non\\
&&~~~~~~~~~
-\,p^{k\a\ad}\Bar{\psi}^{\Bar{a}}_\ad\z^{\Bar{k}}_\a
-{i\over 2}(\pa^{\a\ad}\Bar{B}^{\Bar{k}})\Bar{\psi}^{\Bar{a}}_\ad\rho^k_\a
-\Bar{p}^{\Bar{k}\a\ad}\Bar{\psi}^{\Bar{a}}_\ad\rho^k_\a
\,\Bigg]
\non\\
&&+\,G_{kr\Bar{k}}\Bigg[\,
{i\over 2}(\pa^{\a\ad}B^r)\Bar{\z}^{k}_\ad\z^{\Bar{k}}_\a+
{i\over 2}(\pa^{\a\ad}B^r)\Bar{\rho}^{\Bar{k}}_\ad\rho^k_\a+
{1\over 2}\Bar{S}^{\Bar{k}}_{\Bar{a}}(\pab\Bar{A}^{\Bar{a}})
\Bar{\z}^{k\ad}\Bar{\z}^{r}_\ad+
p^{r\a\ad}\rho^{k}_\a\Bar{\rho}^{\Bar{k}}_\ad+
\Bar{p}^{\Bar{k}\a\ad}\rho^{k}_\a\Bar{\z}^{r}_\ad
\non\\
&&~~~~~~~~~
+\,p^{r\a\ad}\z^{\Bar{k}}_\a\Bar{\z}^{k}_\ad+
{i\over 2}(\pa^{\a\ad}\Bar{B}^{\Bar{k}})\rho^k_\a\Bar{\z}^{r}_\ad+
S^k_a(\pa\,A^a)\rho^{r\a}\z^{\Bar{k}}_\a+
H^r\Bar{\rho}^{\Bar{k}\ad}\Bar{\z}^k_\ad+
{1\over 2}\Bar{H}^{\Bar{k}}\rho^{k\a}\rho^{r}_\a\,\Bigg]
\non\\
&&+\,G_{k\Bar{k}\Bar{r}}\Bigg[\,
{i\over 2}(\pa^{\a\ad}\Bar{B}^{\Bar{r}})\z^{\Bar{k}}_\a\Bar{\z}^{k}_\ad+
{i\over 2}(\pa^{\a\ad}\Bar{B}^{\Bar{r}})\rho^k_\a\Bar{\rho}^{\Bar{k}}_\ad+
{1\over 2}S^{k}_{a}(\pa A^a)
\z^{\Bar{k}\a}\z^{\Bar{r}}_\a
+\Bar{p}^{\Bar{r}\a\ad}\rho^{k}_\a\Bar{\rho}^{\Bar{k}}_\ad+
p^{k\a\ad}\z^{\Bar{r}}_\a\Bar{\rho}^{\Bar{k}}_\ad
\non\\
&&~~~~~~~~~
+\,\Bar{p}^{\Bar{r}\a\ad}\z^{\Bar{k}}_\a\Bar{\z}^{k}_\ad
+{i\over 2}(\pa^{\a\ad}B^{k})\Bar{\rho}^{\Bar{k}}_\ad\z^{\Bar{r}}_\a
+\Bar{S}^{\Bar{k}}_{\Bar{a}}(\pab\,\Bar{A}^{\Bar{a}})\Bar{\rho}^{\Bar{r}\ad}
\Bar{\z}^{k}_\ad+
\Bar{H}^{\Bar{r}}\rho^{k\a}\z^{\Bar{k}}_\a+
{1\over 2}H^{k}\Bar{\rho}^{\Bar{k}\ad}\Bar{\rho}^{\Bar{r}}_\ad
\,\Bigg]
\non\\
&&+\,G_{akr}\Bigg[\,(i\pa^{\a\ad}B^{r})\Bar{\z}^{k}_\ad\psi^{a}_\a+
{i\over 2}(\pa^{\a\ad}A^a)\rho^k_\a\Bar{\z}^{r}_\ad+
S^k_b(\pa A^b)\psi^{a\a}\rho^r_\a+
{1\over 2}F^a\Bar{\z}^{k\ad}\Bar{\z}^{r}_\ad+
p^{r\a\ad}\psi^a_\a\Bar{\z}^{k}_\ad
\,\Bigg]
\non\\
&&+\,G_{\Bar{a}\Bar{k}\Bar{r}}\Bigg[\,
(i\pa^{\a\ad}\Bar{B}^{\Bar{r}})\z^{\Bar{k}}_\a\Bar{\psi}^{\Bar{a}}_\ad+
{i\over 2}(\pa^{\a\ad}\Bar{A}^{\Bar{a}})
\Bar{\rho}^{\Bar{k}}_\ad\z^{\Bar{r}}_\a+
\Bar{S}^{\Bar{k}}_{\Bar{b}}(\pab\, \Bar{A}^{\Bar{b}})
\Bar{\psi}^{\Bar{a}\ad}\Bar{\rho}^{\Bar{r}}_\ad+
{1\over 2}\Bar{F}^{\Bar{a}}\z^{\Bar{k}\a}\z^{\Bar{r}}_\a+
\Bar{p}^{\Bar{r}\a\ad}\z^{\Bar{k}}_\a\Bar{\psi}^{\Bar{a}}_\ad
\,\Bigg]
\non\\
&&+\,G_{krs}\Bigg[\,{1\over 2}H^k\Bar{\z}^{r\ad}\Bar{\z}^s_\ad+
{i\over 2}(\pa^{\a\ad}B^k)\Bar{\z}^{r}_\ad\rho^s_\a+
p^{k\a\ad}\rho^r_\a\Bar{\z}^{s}_\ad+{1\over 2}S^k_a(\pa A^a)\rho^{r\a}\rho^s_\a
\,\Bigg]
\non\\
&&+\,G_{\Bar{k}\Bar{r}\Bar{s}}\Bigg[\,
{1\over 2}\Bar{H}^{\Bar{k}}\z^{\Bar{r}\a}\z^{\Bar{s}}_\a+
{i\over 2}(\pa^{\a\ad}\Bar{B}^{\Bar{k}})\z^{\Bar{r}}_\a\Bar{\rho}^{\Bar{s}}_\ad
+\Bar{p}^{\Bar{k}\a\ad}\z^{\Bar{r}}_\a\Bar{\rho}^{\Bar{s}}_\ad+
{1\over 2}\Bar{S}^{\Bar{k}}_{\Bar{a}}(\pab\, \Bar{A}^{\Bar{a}})
\Bar{\rho}^{\Bar{r}\ad}\Bar{\rho}^{\Bar{s}}_\ad
\,\Bigg]
\non\\
&&+\,{1\over 2}G_{abk}S^k_c(\pa A^c)\psi^{a\a}\psi^b_\a
~+~{1\over 2}G_{\Bar{a}\Bar{b}\Bar{k}}
\Bar{S}^{\Bar{k}}_{\Bar{c}}(\pab\, \Bar{A}^{\Bar{c}})
\Bar{\psi}^{\Bar{a}\ad}\Bar{\psi}^{\Bar{b}}_\ad
\non\\
&&+\,G_{ab\Bar{a}}\Bigg[\,
{i\over 2}(\pa^{\a\ad}A^a)\psi^b_\a\Bar{\psi}^{\Bar{a}}_\ad
+{1\over 2}\Bar{F}^{\Bar{a}}\psi^{a\a}\psi^b_\a\,\Bigg]+
G_{a\Bar{a}\Bar{b}}\Bigg[\,
{i\over 2}(\pa^{\a\ad}\Bar{A}^{\Bar{a}})\Bar{\psi}^{\Bar{b}}_\ad\psi^a_\a
+{1\over 2}F^a\Bar{\psi}^{\Bar{a}\ad}\Bar{\psi}^{\Bar{b}}_\ad\,\Bigg]
\non\\
&&+\,G_{ak\Bar{a}}\Bigg[\,
{i\over 2}(\pa^{\a\ad}B^k)\Bar{\psi}^{\Bar{a}}_\ad\psi^a_\a+
(i\pa^{\a\ad}\Bar{A}^{\Bar{a}})\Bar{\z}^{k}_\ad\psi^{a}_\a+
{i\over 2}(\pa^{\a\ad}A^a)\rho^{k}_\a\Bar{\psi}^{\Bar{a}}_\ad
+F^a\Bar{\z}^{k\ad}\Bar{\psi}^{\Bar{a}}_\ad
\non\\
&&~~~~~~~~~+\,\Bar{F}^{\Bar{a}}\psi^{a\a}\rho^k_\a
+p^{k\a\ad}\psi^a_\a\Bar{\psi}^{\Bar{a}}_\ad\,\Bigg]
\non\\
&&+\,G_{a\Bar{a}\Bar{k}}\Bigg[\,
{i\over 2}(\pa^{\a\ad}\Bar{B}^{\Bar{k}})\psi^a_\a\Bar{\psi}^{\Bar{a}}_\ad+
(i\pa^{\a\ad}A^a)\z^{\Bar{k}}_\a\Bar{\psi}^{\Bar{a}}_\ad+
{i\over 2}(\pa^{\a\ad}\Bar{A}^{\Bar{a}})\Bar{\rho}^{\Bar{k}}_\ad\psi^a_\a
+\Bar{F}^{\Bar{a}}\z^{\Bar{k}\a}\psi^a_\a
\non\\
&&~~~~~~~~~+\,F^a\Bar{\psi}^{\Bar{a}\ad}\Bar{\rho}^{\Bar{k}}_\ad
+\Bar{p}^{\Bar{k}\a\ad}\psi^a_\a\Bar{\psi}^{\Bar{a}}_\ad\,\Bigg]
\non\\
&&+\,G_{ab\Bar{k}}
\Bigg[\,{1\over 2}\Bar{H}^{\Bar{k}}\psi^{a\a}\psi^b_\a+
{1\over 2}(\pa^{\a\ad}A^b)\psi^a_\a\Bar{\rho}^{\Bar{k}}_\ad
\,\Bigg]+
G_{k\Bar{a}\Bar{b}}
\Bigg[\,{1\over 2}H^{k}\Bar{\psi}^{\Bar{a}\ad}\Bar{\psi}^{\Bar{b}}_\ad+
{i\over 2}(\pa^{\a\ad}\Bar{A}^{\Bar{b}})\Bar{\psi}^{\Bar{a}}_\ad\rho^k_\a
\,\Bigg]
\non\\
&&+\,G_{kr\Bar{a}}\Bigg[\,H^r\Bar{\psi}^{\Bar{a}\ad}\Bar{\z}^k_\ad
+{i\over 2}(\pa^{\a\ad}\Bar{A}^{\Bar{a}})\Bar{\z}^k_\ad\rho^r_\a
+{i\over 2}(\pa^{\a\ad}B^r)\Bar{\psi}^{\Bar{a}}_\ad\rho^k_\a
+p^{r\a\ad}\rho^k_\a\Bar{\psi}^{\Bar{a}}_\ad+
{1\over 2}\Bar{F}^{\Bar{a}}\rho^{k\a}\rho^r_\a\,\Bigg]
\non\\
&&+\,G_{a\Bar{k}\Bar{r}}\Bigg[\,\Bar{H}^{\Bar{r}}\psi^{a\a}\z^{\Bar{k}}_\a
+{i\over 2}(\pa^{\a\ad}A^a)\z^{\Bar{k}}_\a\Bar{\rho}^{\Bar{r}}_\ad
+{i\over 2}(\pa^{\a\ad}\Bar{B}^{\Bar{r}})\psi^a_\a\Bar{\rho}^{\Bar{k}}_\ad
+\Bar{p}^{\Bar{r}\a\ad}\psi^a_\a\Bar{\rho}^{\Bar{k}}_\ad+
{1\over 2}F^a\Bar{\rho}^{\Bar{k}\ad}\Bar{\rho}^{\Bar{r}}_\ad
\,\Bigg]
\non\\
&&+\,{1\over 4}G_{ab\Bar{a}\Bar{b}}\,\psi^{a\a}\psi^b_\a
\,\Bar{\psi}^{\Bar{a}\ad}\Bar{\psi}^{\Bar{b}}_\ad
~+~{1\over 2}G_{ak\Bar{a}\Bar{b}}\,\psi^{a\a}\rho^k_\a\,
\Bar{\psi}^{\Bar{a}\ad}\Bar{\psi}^{\Bar{b}}_\ad
~+~{1\over 2}G_{ab\Bar{a}\Bar{k}}\,\psi^{a\a}\psi^b_\a\,
\Bar{\psi}^{\Bar{a}\ad}\Bar{\rho}^{\Bar{k}}_\ad
\non\\
&&+\,G_{ak\Bar{a}\Bar{k}}\,\Bigg(\psi^{a\a}\z^{\Bar{k}}_\a\,
\Bar{\z}^{k\ad}\Bar{\psi}^{\Bar{a}}_\ad+
\psi^{a\a}\rho^k_\a\,
\Bar{\psi}^{\Bar{a}\ad}\Bar{\rho}^{\Bar{k}}_\ad\Bigg)~+~
{1\over 2}G_{abk\Bar{a}}\,\psi^{a\a}\psi^{b}_\a\,
\Bar{\psi}^{\Bar{a}\ad}\Bar{\z}^k_\ad
\non\\
&&
+\,{1\over 2}G_{a\Bar{a}\Bar{b}\Bar{k}}\,\psi^{a\a}\z^{\Bar{k}}_\a\,
\Bar{\psi}^{\Bar{a}\ad}\Bar{\psi}^{\Bar{b}}_\ad~+
~G_{a\Bar{a}\Bar{k}\Bar{r}}\,\psi^{a\a}\z^{\Bar{k}}_\a\,
\Bar{\psi}^{\Bar{a}\ad}\Bar{\rho}^{\Bar{r}}_\ad~+
~G_{akr\Bar{a}}\,\psi^{a\a}\rho^k_\a\,\Bar{\z}^{r\ad}\Bar{\psi}^{\Bar{a}}_\ad
\non\\
&&
+\,{1\over 4}G_{abkr}\,\psi^{a\a}\psi^b_\a\,\Bar{\z}^{k\ad}\Bar{\z}^r_\ad
~+~{1\over 4}G_{\Bar{a}\Bar{b}\Bar{k}\Bar{r}}\,\z^{\Bar{k}\a}\z^{\Bar{r}}_\a\,
\Bar{\psi}^{\Bar{a}\ad}\Bar{\psi}^{\Bar{b}}_\ad~
+\,{1\over 2}G_{akrs}\,\psi^{a\a}\rho^k_\a\,\Bar{\z}^{r\ad}\Bar{\z}^s_\ad
\non\\
&&+\,
{1\over 2}G_{\Bar{a}\Bar{k}\Bar{r}\Bar{s}}\,\z^{\Bar{k}\a}\z^{\Bar{r}}_\a\,
\Bar{\psi}^{\Bar{a}\ad}\Bar{\rho}^{\Bar{s}}_\ad~+
{1\over 4}G_{krst}\,\rho^{k\a}\rho^r_\a\,\Bar{\z}^{s\ad}\Bar{\z}^t_\ad~
+{1\over 4}G_{\Bar{k}\Bar{r}\Bar{s}\Bar{t}}\,\z^{\Bar{k}\a}\z^{\Bar{r}}_\a\,
\Bar{\rho}^{\Bar{s}\ad}\Bar{\rho}^{\Bar{t}}_\ad
\non\\
&&+\,G_{akr\Bar{k}}\Bigg(
\psi^{a\a}\rho^k_\a\,\Bar{\z}^{r\ad}\Bar{\rho}^{\Bar{k}}_\ad
+{1\over 2}\psi^{a\a}\z^{\Bar{k}}_\a\,\Bar{\z}^{k\ad}\Bar{\z}^{r}_\ad\Bigg)~+
~{1\over 2}G_{abk\Bar{k}}\,\psi^{a\a}\psi^b_\a\,
\Bar{\z}^{k\ad}\Bar{\rho}^{\Bar{k}}_\ad
\non\\
&&+\,{1\over 2}G_{k\Bar{a}\Bar{b}\Bar{k}}\,\rho^{k\a}\z^{\Bar{k}}_\a
\,\Bar{\psi}^{\Bar{a}\ad}\Bar{\psi}^{\Bar{b}}_\ad~+~
G_{k\Bar{a}\Bar{k}\Bar{r}}\Bigg(
\rho^{k\a}\z^{\Bar{k}}_\a\,\Bar{\psi}^{\Bar{a}\ad}\Bar{\rho}^{\Bar{r}}_\ad
+{1\over 2}\z^{\Bar{k}\a}\z^{\Bar{r}}_\a\,
\Bar{\psi}^{\Bar{a}\ad}\Bar{\z}^{k}_\ad\Bigg)
\non\\
&&+\,G_{ak\Bar{k}\Bar{r}}\Bigg(\psi^{a\a}\z^{\Bar{k}}_\a\,
\Bar{\z}^{k\ad}\Bar{\rho}^{\Bar{r}}_\ad
+{1\over 2}\psi^{a\a}\rho^k_\a\,\Bar{\rho}^{\Bar{k}\ad}\Bar{\rho}^{\Bar{r}}_\ad
\Bigg)~+
~{1\over 4}G_{ab\Bar{k}\Bar{r}}\,\psi^{a\a}\psi^{b}_\a\,
\Bar{\rho}^{\Bar{k}\ad}\Bar{\rho}^{\Bar{r}}_\ad
\non\\
&&+\,{1\over 4}G_{kr\Bar{a}\Bar{b}}\,\rho^{k\a}\rho^{r}_\a\,
\Bar{\psi}^{\Bar{a}\ad}\Bar{\psi}^{\Bar{b}}_\ad~+
~G_{kr\Bar{a}\Bar{k}}\Bigg(\rho^{k\a}\z^{\Bar{k}}_\a\,
\Bar{\z}^{r\ad}\Bar{\psi}^{\Bar{a}}_\ad
+{1\over 2}\rho^{k\a}\rho^r_\a\,\Bar{\psi}^{\Bar{a}\ad}\Bar{\rho}^{\Bar{k}}_\ad
\Bigg)
\non\\
&&
+\,{1\over 2}G_{krs\Bar{k}}\Bigg(\rho^{k\a}\z^{\Bar{k}}_\a\,
\Bar{\z}^{r\ad}\Bar{\z}^{s}_\ad+
\rho^{k\a}\rho^{r}_\a\,\Bar{\z}^{s\ad}\Bar{\rho}^{\Bar{k}}_\ad
\Bigg)~+
~{1\over 2}G_{krs\Bar{a}}\,\rho^{k\a}\rho^r_\a\,
\Bar{\z}^{s\ad}\Bar{\psi}^{\Bar{a}}_\ad
\non\\
&&+\,{1\over 2}G_{a\Bar{k}\Bar{r}\Bar{s}}\,\psi^{a\a}\z^{\Bar{k}}_\a\,
\Bar{\rho}^{\Bar{r}\ad}\Bar{\rho}^{\Bar{s}}_\ad~+
~{1\over 2}G_{k\Bar{k}\Bar{r}\Bar{s}}\Bigg(
\z^{\Bar{k}\a}\z^{\Bar{r}}_\a\,\Bar{\z}^{k\ad}\Bar{\rho}^{\Bar{s}}_\ad+
\rho^{k\a}\z^{\Bar{k}}_\a\,\Bar{\rho}^{\Bar{r}\ad}\Bar{\rho}^{\Bar{s}}_\ad
\Bigg)
\non\\
&&+\,G_{kr\Bar{k}\Bar{r}}\Bigg(
\rho^{k\a}\z^{\Bar{k}}_\a\,\Bar{\z}^{r\ad}\Bar{\rho}^{\Bar{r}}_\ad
+{1\over 4}\rho^{k\a}\rho^{r}_\a\,
\Bar{\rho}^{\Bar{k}\ad}\Bar{\rho}^{\Bar{r}}_\ad
+{1\over 4}\z^{\Bar{k}\a}\z^{\Bar{r}}_\a\,\Bar{\z}^{k\ad}\Bar{\z}^{r}_\ad
\Bigg)~~~.
\label{ScomponentsCNM6D}
\eea

\end{document}